\documentclass[manuscript]{aastex}
\usepackage{graphicx}
\usepackage{ulem}

\newcommand{\kms}[1]{$#1$~km~s$^{-1}$}
\newcommand{\amm}[1]{$#1$~A~m$^{-2}$}

\begin{document}

\title{The Formation of a Magnetic Channel by the Emergence of
Current-carrying Magnetic Fields}
\author{Eun-Kyung Lim\altaffilmark{1}, Jongchul Chae\altaffilmark{1,2},
Ju Jing\altaffilmark{3}, Haimin Wang\altaffilmark{3} and Thomas
Wiegelmann\altaffilmark{4}} \email{eklim@astro.snu.ac.kr,
jcchae@snu.ac.kr} \altaffiltext{1}{Astronomy Program, Department of
Physics and Astronomy, Seoul National University, Seoul 151-742,
Republic of Korea} \altaffiltext{2}{Big Bear Solar Observatory, New
Jersey Institute of Technology, 40386 North Shore Lane, Big Bear City,
CA 92314-9672, USA} \altaffiltext{3}{Space Weather Research
Laboratory, New Jersey Institute of Technology, Newark, NJ 07102, USA}
\altaffiltext{4}{Max-Planck Institute f\"ur Sonnensystemforschung,
Max-Planck, Strasse 2, 37191, Katlenburg-Lindau, Germany}


\begin{abstract}
A magnetic channel -- a series of polarity reversals separating
elongated flux threads with opposite polarities -- may be a
manifestation of  a highly non-potential magnetic configuration in
active regions. To understand its formation we have carried out a
detailed analysis of the magnetic channel in AR 10930 using data taken
by the Solar Optical Telescope /\textit{Hinode}. As a result, we found
upflows ($-0.5$ to \kms{-1.0}) and downflows ($+1.5$ to \kms{+2.0})
inside and at both tips of the thread respectively, and a pair of
strong vertical currents of opposite polarity along the channel.
Moreover, our analysis of the nonlinear force-free fields constructed
from the photospheric magnetic field indicates that the current
density in the lower corona may have gradually increased as a result
of the continuous emergence of the highly sheared flux along the
channel. With these results, we suggest that the magnetic channel
originates from the emergence of the twisted flux tube that has formed
below the surface before the emergence.
\end{abstract}


\section{Introduction}
Flares usually occur in regions of highly non-potential magnetic field
that can contain large amount of free magnetic energy
\citep{WanJ96,Sch09}. The degree of magnetic non-potentiality has been
empirically measured by different kinds of observational parameters
such as magnetic gradient across the polarity inversion lines (PIL)
\citep{Wang06}, magnetic shear at PILs \citep{Hag84,Hag90} and flux
emergence at the flaring site \citep{Ish98,Sch09}. Most of highly
non-potential magnetic fields have complex configurations.

\citet{Zir93} observed a complex magnetic field configuration called
magnetic channel in an active $\delta$-type sunspot group that
produced major flares. From the vector videomagnetograms taken at the
Big Bear Solar Observatory, they found a series of alternating
opposite polarities near the main PIL of the active region where major
flares occurred about one day after. Those elongated opposite
polarities were separated by channels along which the strong
transverse magnetic fields were detected. Even though they did not
show the direct connection between the magnetic channel and the flare
occurrence, the appearance of such a sheared structure at the flaring
region itself is interesting in the viewpoint of the magnetic
non-potentiality of the flaring region.

The characteristics of a magnetic channel were recently studied by
\citet{wang08} with data from the Solar Optical Telescope
(SOT)/\textit{Hinode}. This channel was observed between two spots of
opposite polarity in AR 10930. By analyzing both vector magnetograms
and three-dimensional coronal field obtained from the nonlinear
force-free field (NLFFF) extrapolation, they quantified magnetic
properties of the channel and confirmed that the channel {was} a
highly sheared {and non-potential} structure. {For instance, they
measured the distribution of shear angles in the channel area using
vector magnetograms and showed that the median value was
63$^{\circ}$.9.} They also suggested a schematic model to explain the
formation mechanism of the channel. According to this model, a series
of small bipoles become squeezed and elongated once they emerge
because of the compactness of the background field, leading to the
formation of the magnetic channel. Unfortunately, however, since they
analyzed only a snapshot vector magnetogram, the study was not
sufficient to show the formation process of the magnetic channel.
Their analysis focused on the single set of magnetic data taken at 14
UT on December 13 when the magnetic channel was already well
developed.

We aim to systematically investigate the formation process of the
channel structure in AR 10930. Since a time {sequential
Spectropolarimeter (SP) data from SOT} are available at the early
phase of the channel formation, more detailed analysis could be
performed at this time period. As \citet{wang08} mentioned, the
channel is likely to have been formed due to flux emergence at the
flaring region. Note that the flux emergence has been regarded as one
of the important processes that increase the magnetic non-potentiality
\citep{Kur87,Tan91,Ish98,Kur02,Bro03,Sch09}. In addition, it seems
that the formation of the magnetic channel could be related to the
flare that occurred one day later, on December 13. If we investigate
how such a sheared structure developed prior to the flare occurrence,
we will be able to better understand how free magnetic energy was
accumulated and how the destabilization of the magnetic field took
place in AR 10930. Like the active region studied by \citet{Zir93}, AR
10930 also shows the $\delta$-configuration and produced a number of
flares including three X-class ones during its first disk passage. Due
to its high degree of activity, specifically the X-class flare on 2006
December 13, many flare-related studies have been performed with AR
10930 using data from SOT/\textit{Hinode}
\citep{Kub07,Guo08,Sch08,Mag08}.

Specifically, we examine the temporal evolution of the photospheric
vector magnetic fields around the formation site of the channel. We
also probe the three-dimensional structures of both magnetic field and
electric current in the corona above the magnetic channel by applying
the NLFFF extrapolation.

The content of this paper is organized as follows. In Section 2, we
describe the data and the analysis methods. Then the observational
findings and their significance are presented in Section 3 that is
divided into two parts. In Section 3.1, the photospheric evolution of
the channel structure examined by analyzing Filtergraph (FG) and SP
data is presented. The description of the temporal change in the
Doppler velocity and the vector magnetic field is also included. In
Section 3.2, the characteristics of the coronal magnetic field and the
coronal current density deduced from the NLFFF extrapolation are
presented. Finally, we summarize the key results and discuss their
physical implications in Section 4.

\section{Observations and Data Analysis}
\subsection{Data description}
AR 10930 was observed from 2006 December 5 to 17 in the southern
hemisphere near the solar equator. This active region is one of the
highly flare-productive active regions well observed by
\textit{Hinode}, the Japanese satellite launched on 2006 September 22.
It showed a significant rotation of sunspots during its first disk
passage \citep{Min09}, and several flares occurred meanwhile including
the X-3.4 flare of 2006 December 13. Figure~\ref{fov_chn} shows the
line-of-sight (LOS) magnetogram obtained from SP data using the
center-of-gravity (COG) method \citep{Rees79,Uit03,Chae07}. The COG
method could be a good choice when interested in the LOS component of
magnetic field only, since it takes shorter inversion time and does
not require any radiative transfer model. The data were taken about
1.5 days before the X-3.4 flare. Between the two spots of opposite
polarity, there are several faint threads elongated in the east-west
direction. In specific, the thin structure of negative polarity
indicated by the arrow is the object of our main interest. We
investigate its formation and temporal evolution for about one day.

The SOT on board \textit{Hinode} provides SP data that consist of
full Stokes profiles of the two Fe lines at 6301.5 and 6302.5 {\AA}
taken using the slit of 0.16\arcsec\ by 164\arcsec. We have examined
7 sets of SP data that were taken during the period of our interest
from 17 UT on December 11 to 20 UT on the 12th. Every set covered
the same field of view of about 280\arcsec\ by 160\arcsec, which is
large enough to cover the entire active region. One set was taken
with the normal mode at 11 UT on December 12, which is characterized
by the high spatial sampling of 0.16\arcsec, and a long scan time of
about 160 minutes. The others were taken with the fast mode, with a
coarser spatial sampling of 0.32\arcsec\ and the shorter scan time
of about 60 minutes. Although the coarser spatial sampling of the
fast-mode scan costs the spatial resolution, its shorter scan time
makes it proper for this study considering the fast change of the
channel structure. The SOT also produced Stokes $I$ and $V$
filtergrams (FG) with a higher time cadence of a few minutes. These
data were useful in the study of the morphological change of
magnetic structures.

\subsection{Data analysis}
In order to measure the LOS velocity of the active region, we have
applied the COG method to Stokes $I$ profiles at all the pixels in the
channel site. The COG wavelength of the averaged $I$ profile of the
quiet Sun was taken as the reference. It has been demonstrated by
\citet{Uit03} that the COG method accurately derives the LOS velocity
even in the case of asymmetric lines. This property makes it suitable
to apply the method to our analysis since a number of the line
profiles in the region of our interest near the PIL are found to be
asymmetric probably because of the complexity of magnetic and velocity
structures there.

The vector magnetic fields were retrieved by using the conventional
Stokes inversion method that fits the observed line profiles assuming
the Milne--Eddington (ME) atmosphere. The $180^{\circ}$ ambiguity was
resolved by applying the Uniform Shear Method (USM) introduced by
\citet{Moon03}. These authors showed that the method successfully
removed the spatial discontinuities of the transverse fields of active
regions containing highly sheared regions. \citet{Moon07} also adopted
USM in resolving the azimuth ambiguity of AR 10930, the same active
region as the one we are studying, and showed that the method
successfully worked on all areas.

\begin{figure}[tb]
\begin{center}
    \includegraphics[width=0.5\textwidth]{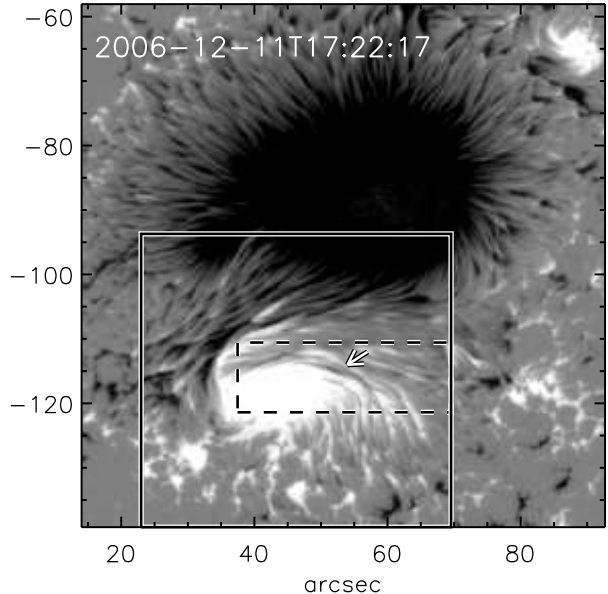}\\
    \caption{LOS magnetogram of AR 10930.
    The arrow indicates the location of the negative flux thread which is
    our object of interest. The solid rectangle represents the FOV for both
    the preprocessing and the NLFFF extrapolation, and the dashed rectangle
    indicates the region
    that we cut for the image display in Figures~\ref{dopp},
    \ref{vectogram}, \ref{incli}, and \ref{current}.}\label{fov_chn}
\end{center}
\end{figure}

{After correcting the projection effect,} the small part of the active
region that contains the channel structure of our interest was chosen
from each of vector magnetograms {for the bottom boundary condition of
the NLFFF extrapolation. In the optimization code devised by
\citet{Wie04}, the lateral and the top boundaries are assumed to be
potential field. If the boundaries are far from the active region, the
effect of the lateral boundaries on the extrapolated field could be
negligible. In our case, the solid rectangle in Figure~\ref{fov_chn}
shows the FOV for the NLFFF extrapolation. The magnetic structure of
our interest is relatively small and far from lateral boundaries, so
that field lines of this structure are likely to be closed within the
computational box. Therefore, we expect that the effect of the lateral
boundary condition on the computed fields would not be significant.}
Since the magnetic channel is fine-scale, we maintained the original
spatial sampling of 0\arcsec.32 for the extrapolation. Instead, we
limited the FOV for the extrapolation to about 50\arcsec\ by
50\arcsec\ to reduce both the computational time and the size of data.

\begin{figure}[tb]
\begin{center}
    \plottwo{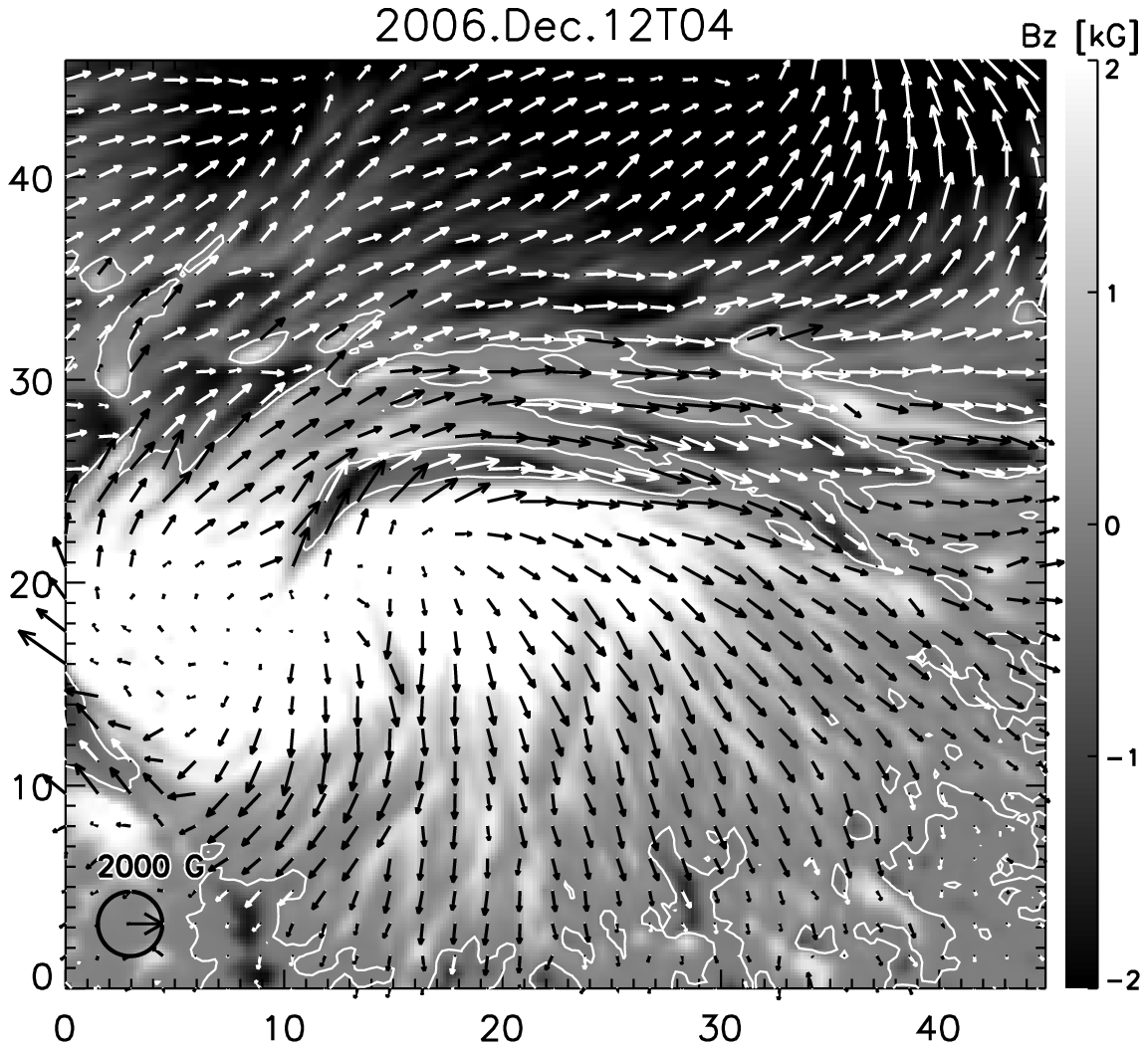}{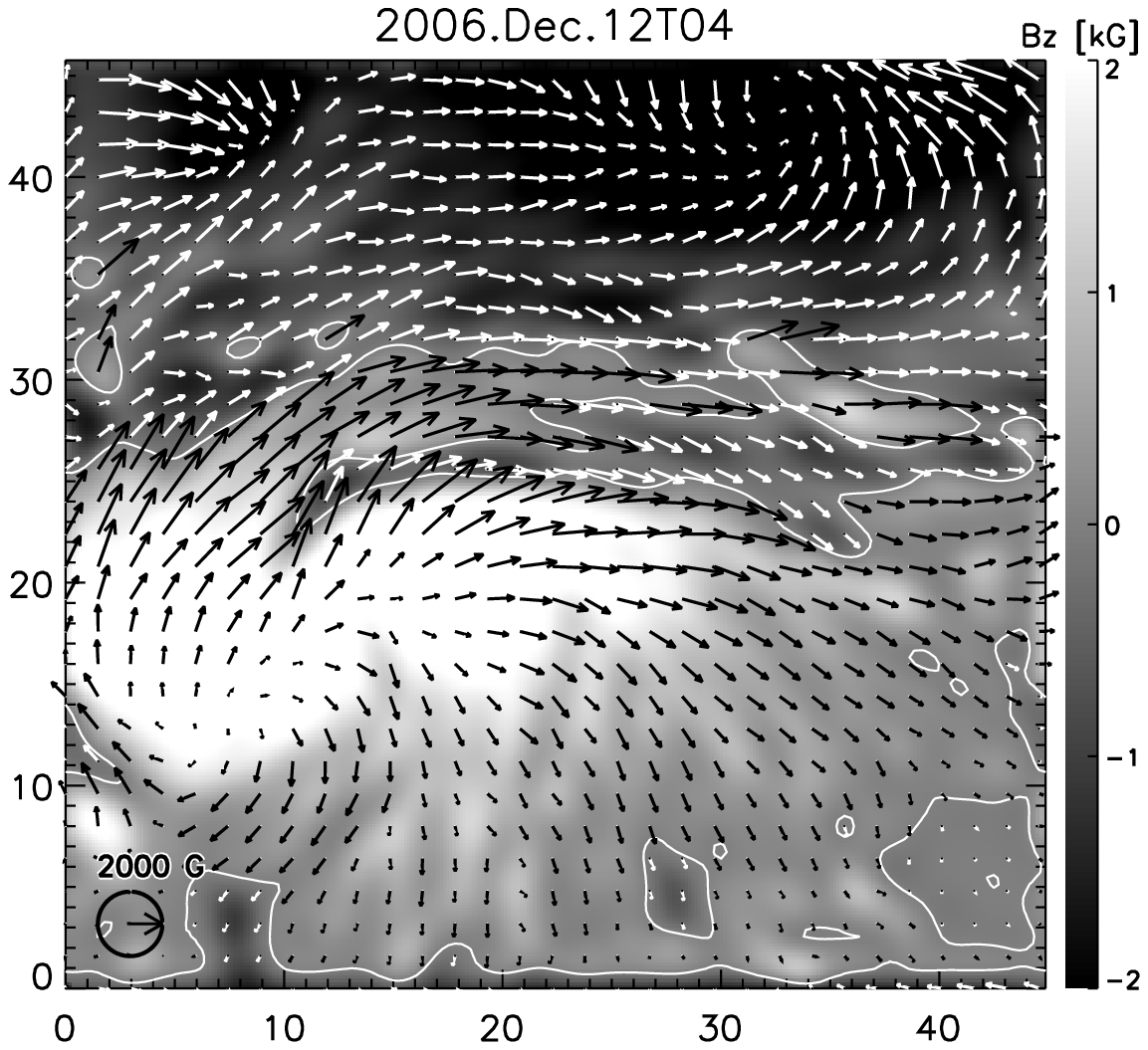}
    \caption{Vector magnetograms of AR 10930 without preprocessing (left)
    and with preprocessing ({right}).}\label{prep}
\end{center}
\end{figure}

Before the NLFFF extrapolation, all data were preprocessed following
the procedure developed by \citet{Wie06} in order to make the
photospheric magnetic field close to force-free. {Preprocessing alters
the vector magnetic field so that the bottom boundary satisfies the
force-free and torque-free condition \citep{Wie06} and also represents
the chromospheric force-free condition better \citep{Jing10}.
Figure~\ref{prep} shows vector magnetograms before and after the
preprocessing. Although fine-scale structures have been smoothed and
their boundaries have been broadened after preprocessing, the channel
structure is still well recognized. The directions and strength of
transverse fields have also been slightly altered. The direction of
transverse fields changed to be less parallel to the neutral line of
the channel structure indicating that the magnetic shear in the
channel region may have been reduced after preprocessing. The
relatively large change of transverse fields near the negative umbral
region may be due to the boundary effect. Since this region is over
15~Mm away from the channel structure, it is expected that magnetic
fields with lower heights, roughly half of 15~Mm, may not be affected
significantly by the boundary effect.}


\section{Results}

\subsection{Photospheric magnetic and velocity fields}
\begin{figure}[tb]
\begin{center}
    \includegraphics[width=0.7\textwidth]{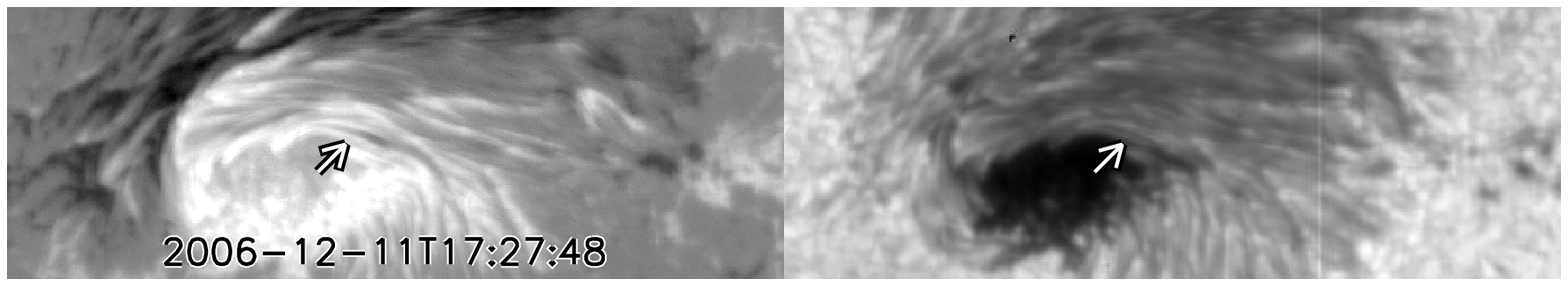}\\\,\,
    \includegraphics[width=0.7\textwidth]{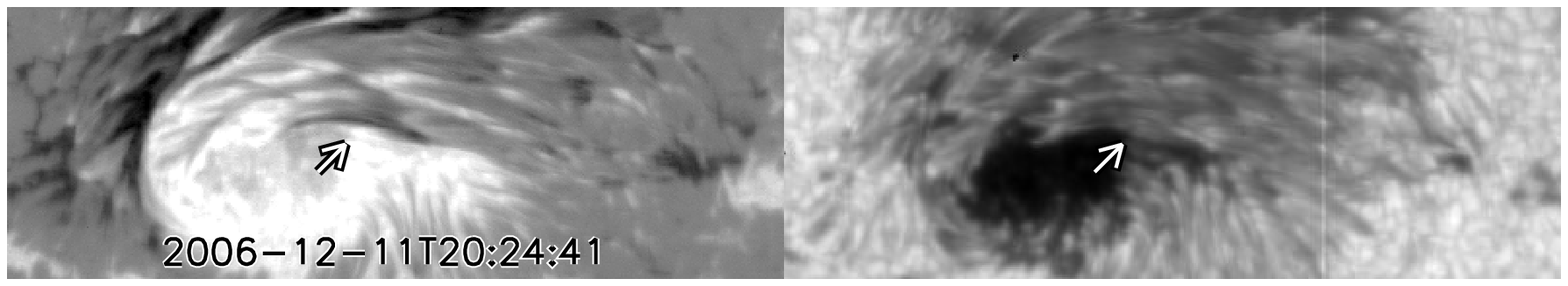}\\\,\,
    \includegraphics[width=0.7\textwidth]{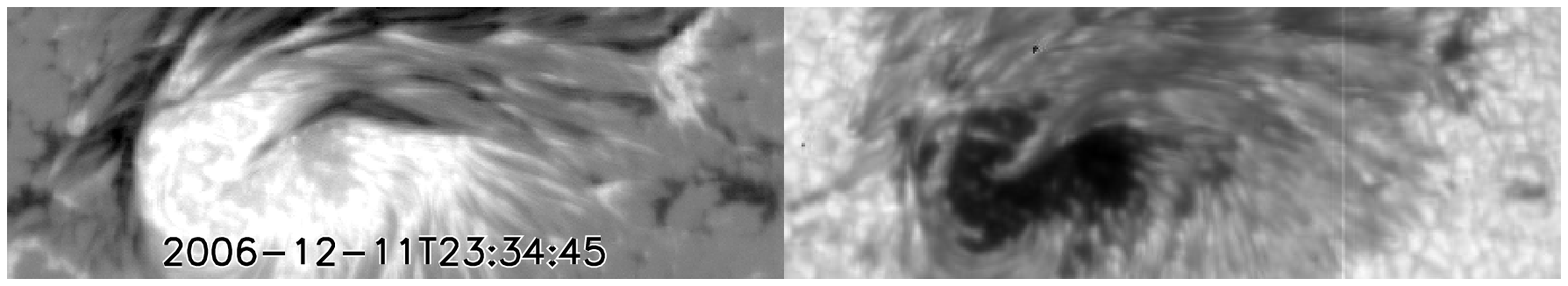}\\\,\,
    \includegraphics[width=0.7\textwidth]{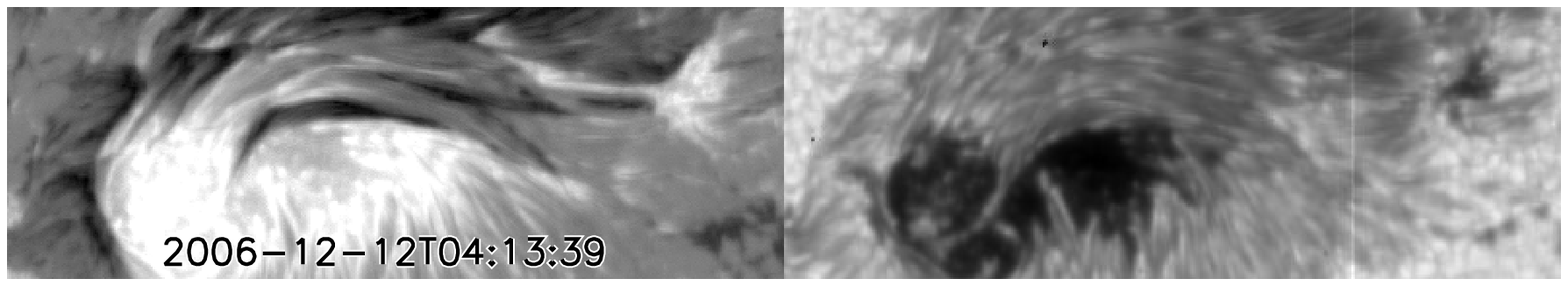}\\\,\,
    \includegraphics[width=0.7\textwidth]{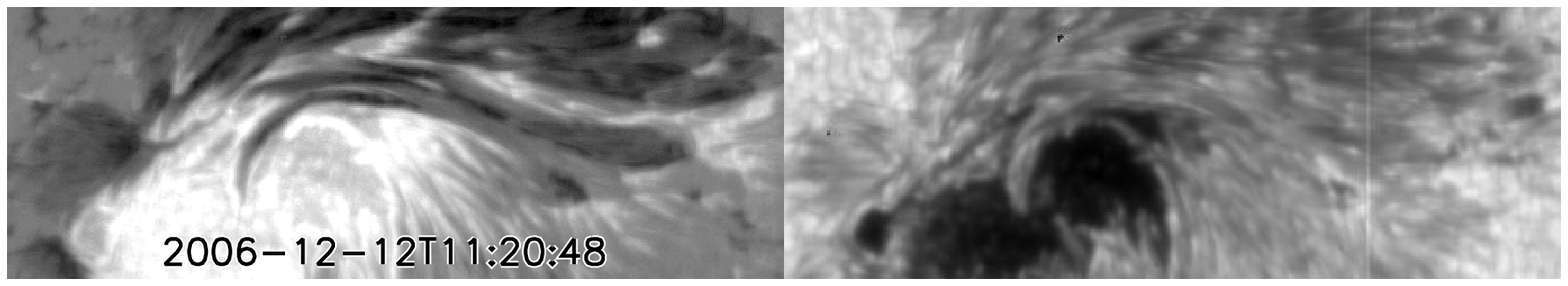}\\\,\,
    \includegraphics[width=0.7\textwidth]{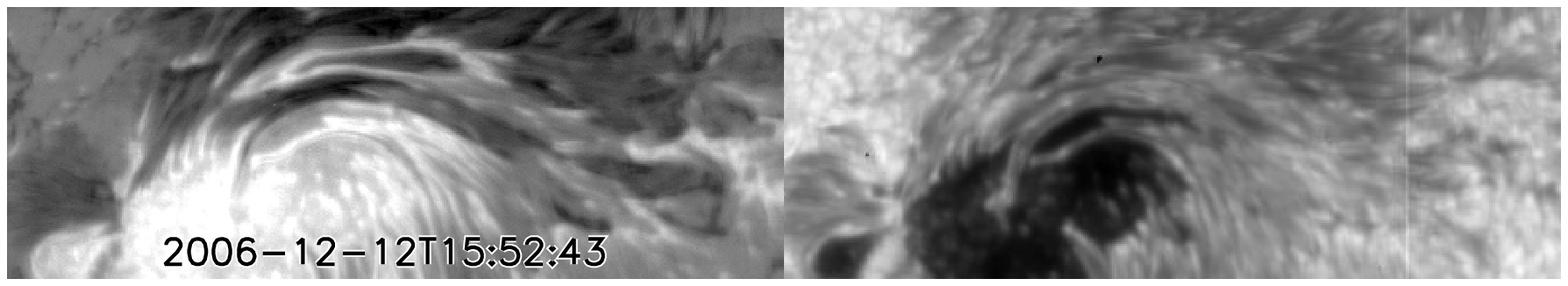}\\\,\,
    \includegraphics[width=0.7\textwidth]{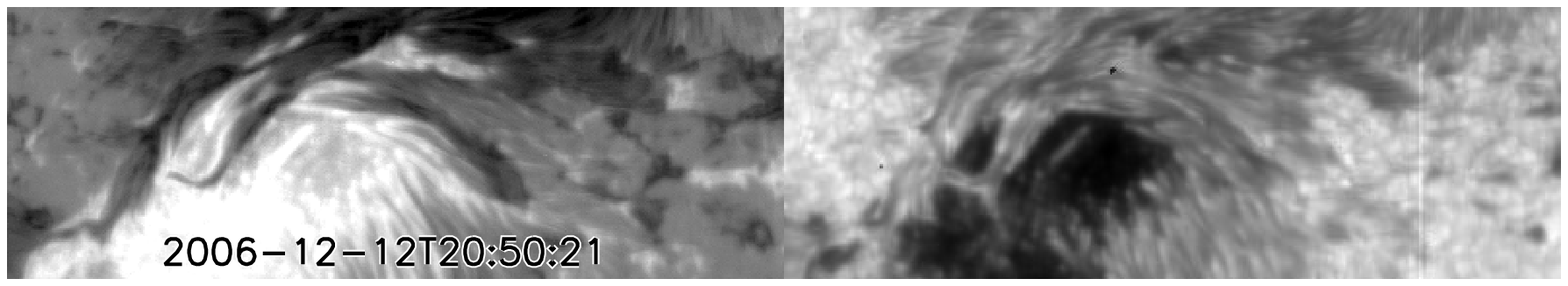}\\\,\,
    \includegraphics[width=0.7\textwidth]{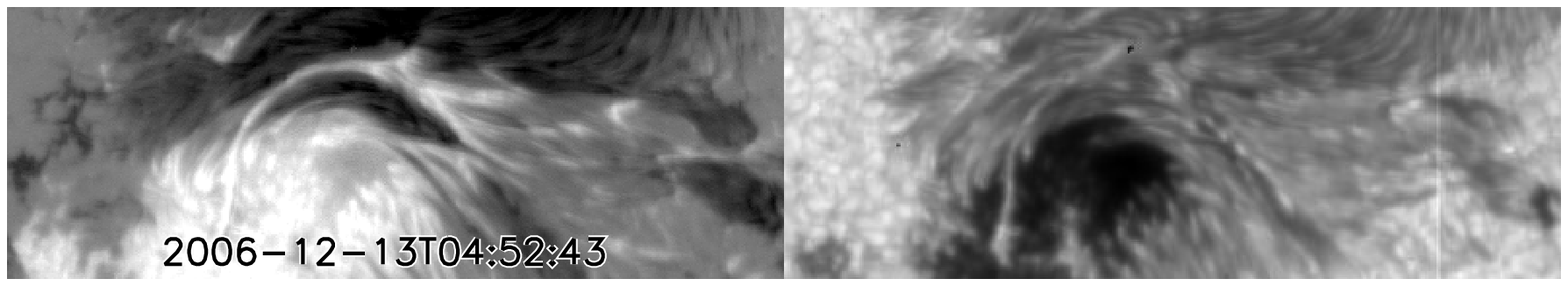}\\\,\,
    \caption{FG $V$/$I$ maps ({left}) and the corresponding FG $I$ maps ({right}).
    Images in the same row are taken at the same time and show the same FOV
    slightly larger than that of the white box in Figure~\ref{fov_chn}.
    Arrows in images in the first and the second row are pointing the
    location of the negative flux thread of our interest that is composing
    the magnetic channel.}\label{temp}
\end{center}
\end{figure}

Figure~\ref{temp} presents the time series of FG $V/I$ and $I$ maps
showing the process of the formation and evolution of the magnetic
channel. The most notable in the $V/I$ maps is the gradual appearance
and the fast growth of a thread-like structure at the edge of the
positive spot near the main PIL. As shall be shown later in this
section, this process of appearance and growth physically represents
the emergence of magnetic flux from below. The formation stage of this
structure is well recorded in the three magnetograms taken on December
11. The first magnetogram taken at 17:27 UT on December 11 just before
the formation of the channel shows that the region  is dominated by
the magnetic field of positive polarity with a strength of about 3000
G. However, at the elongated area pointed by an arrow, the magnetic
field is weak with a strength of a few hundreds of Gauss only. This is
the very location where the negative flux thread is to appear. At
20:24 UT on December 11, the polarity of the magnetic field in the
area changes to negative and the thread becomes clearly visible on the
surface. Although it represents the very early phase of emergence, the
shape of the negative flux is quite elongated in contrast to the usual
emergence of a bipole.

This negative flux thread seen in magnetograms corresponds to a dark
penumbral filament in $I$ maps. This penumbral filament can be
identified at the elongated area of weak field even at 17:27 UT on
the 11th, before the appearance of the negative flux thread. It gets
longer and more prominent as the negative flux thread shows up and
increases in length and field strength. Supposing that the magnetic
field is along the penumbral filament, the detection of the
penumbral filament before the appearance of the magnetic flux thread
suggests the existence of the sheared fields beneath the
photosphere.

Figure~\ref{temp} presents a nice example of high-resolution
observation showing how a magnetic channel is fully developed. After
emergence, the magnetic configuration of the first flux thread becomes
complex. A few hours after its appearance, at about 23:34 UT on
December 11, another thread of negative polarity appears between the
first one and the PIL. This grows faster than the first one and
significantly changes its shape within only 16 hr. Due to this kind of
sequential emergence of negative flux threads, the polarity of the LOS
magnetic field reverses five times across threads at 15:52 UT on
December 12, resulting in a well-developed magnetic channel structure.

\begin{figure}[tb]
\begin{center}
    \includegraphics[width=0.4\textwidth]{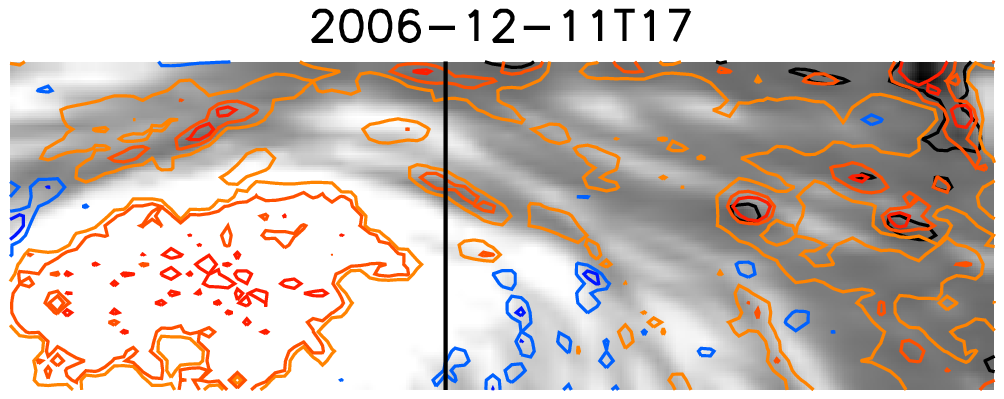}
    \includegraphics[width=0.4\textwidth]{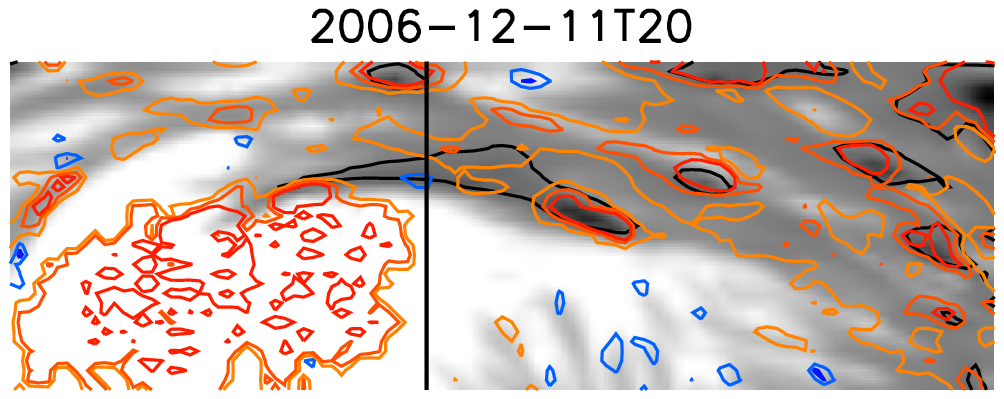}\\
    \includegraphics[width=0.4\textwidth]{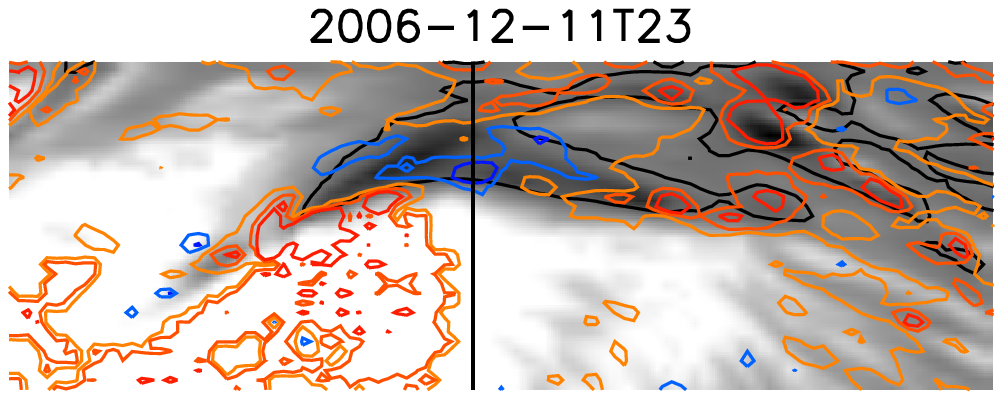}
    \includegraphics[width=0.4\textwidth]{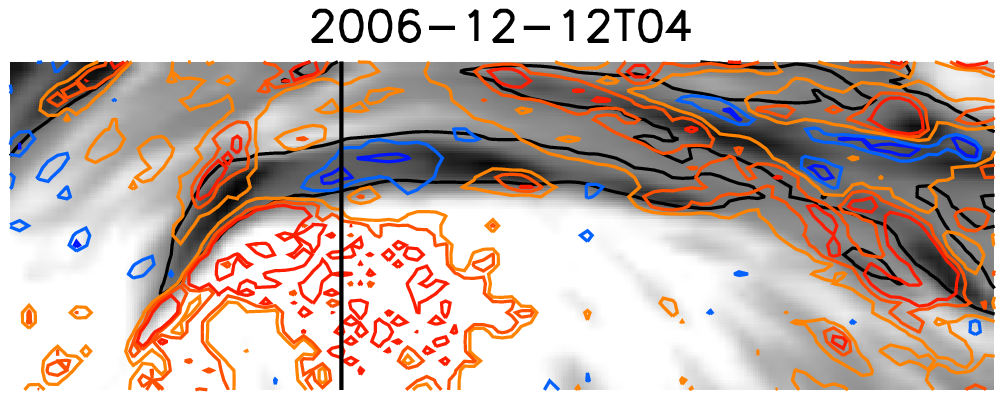}\\
    \includegraphics[width=0.4\textwidth]{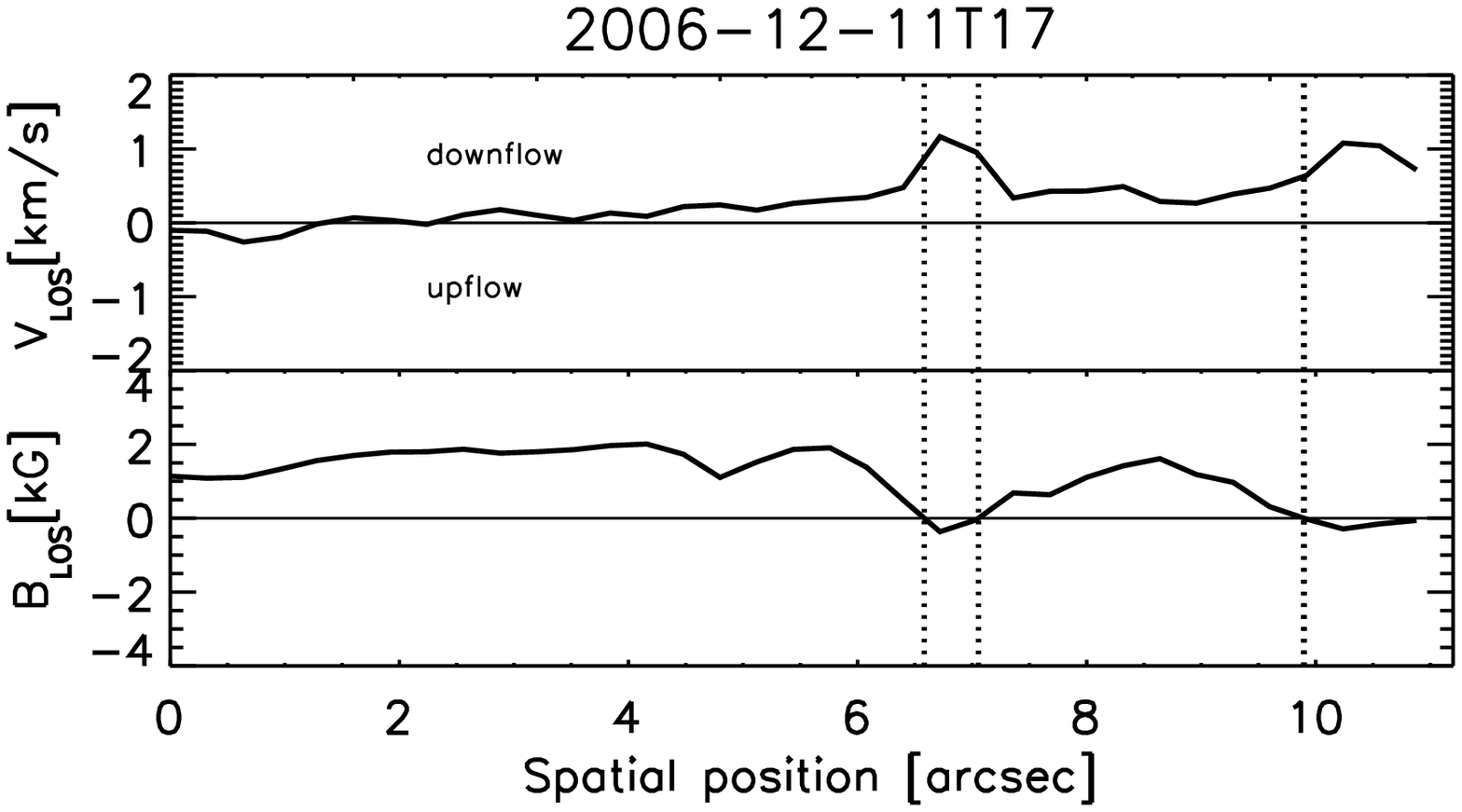}
    \includegraphics[width=0.4\textwidth]{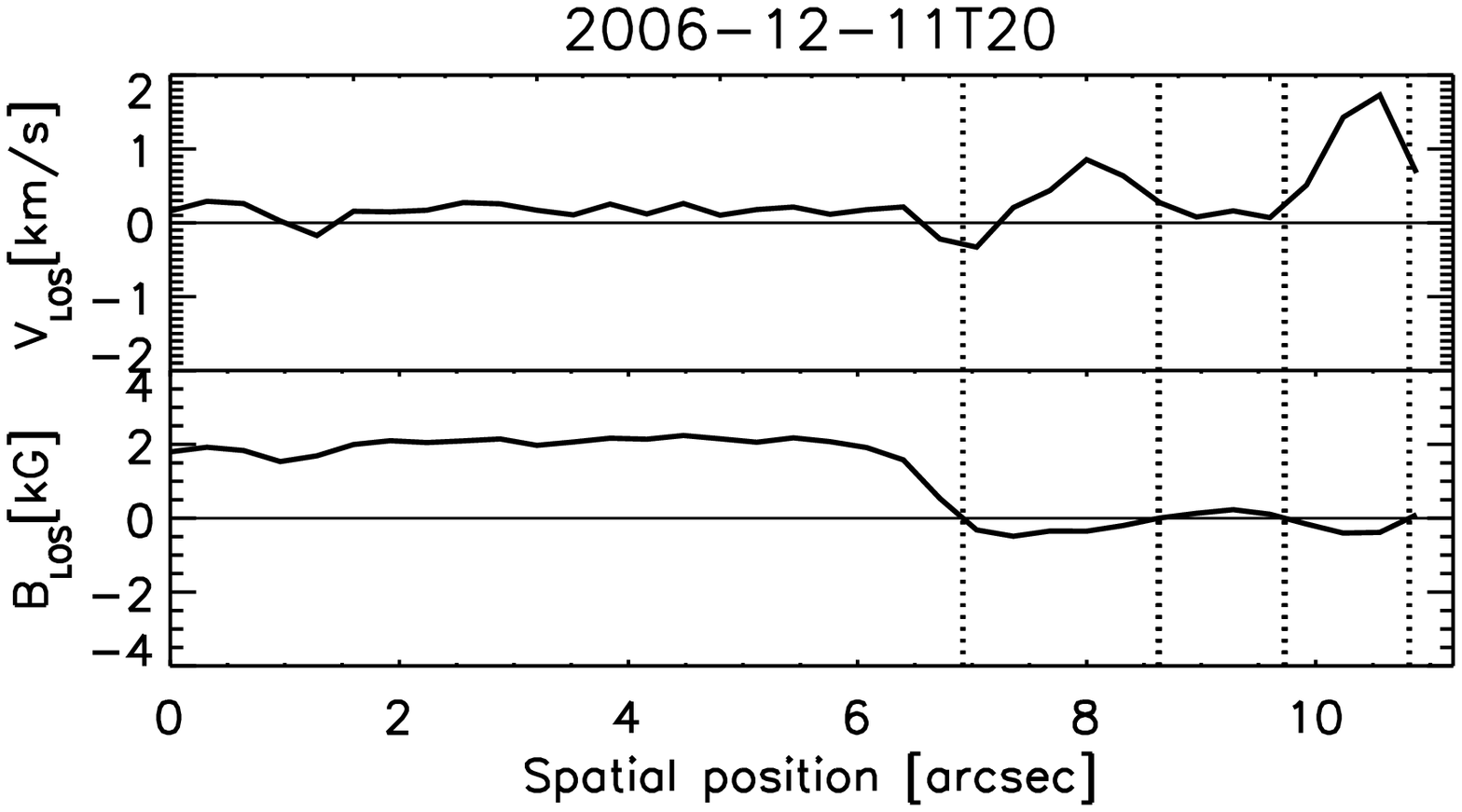}\\
    \includegraphics[width=0.4\textwidth]{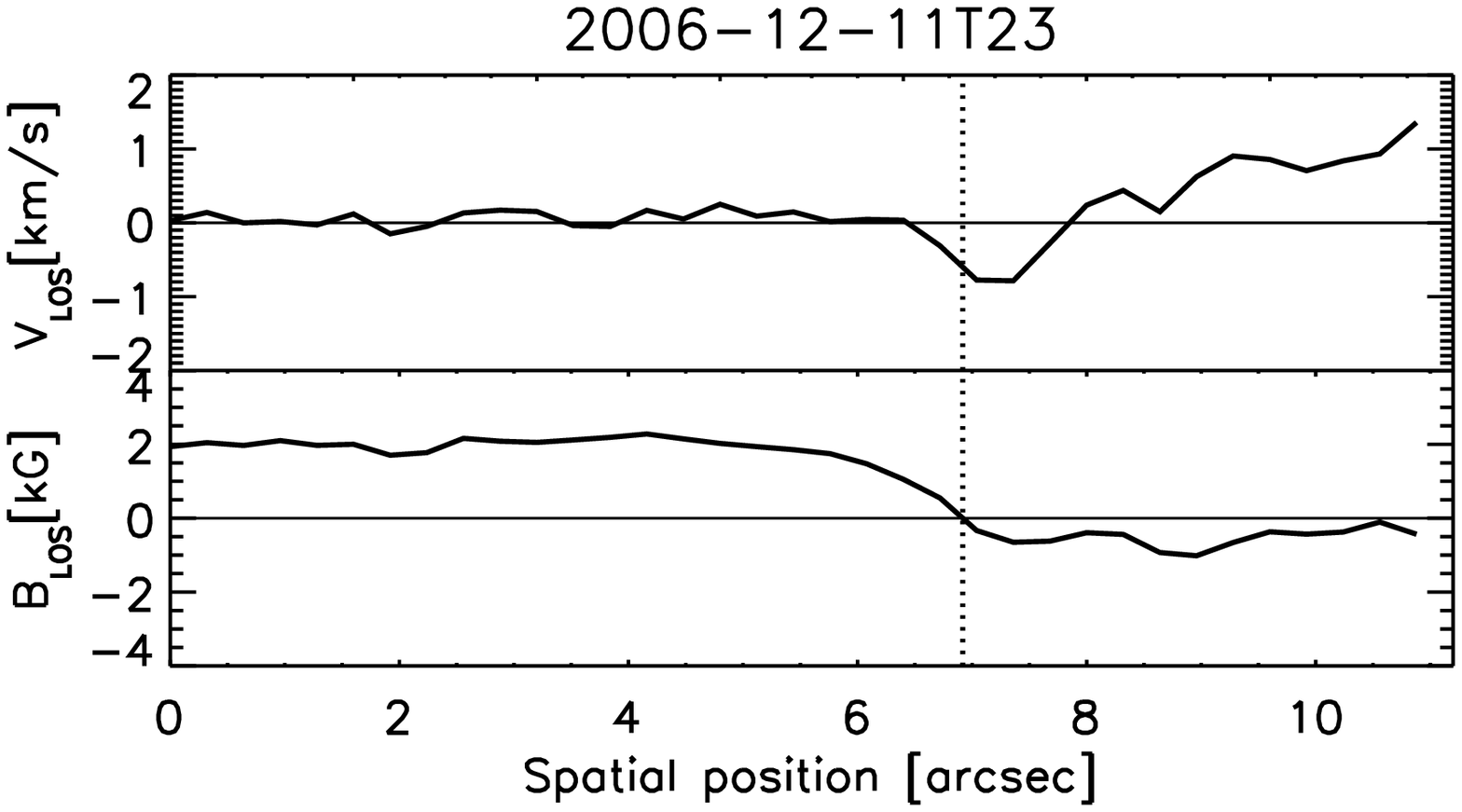}
    \includegraphics[width=0.4\textwidth]{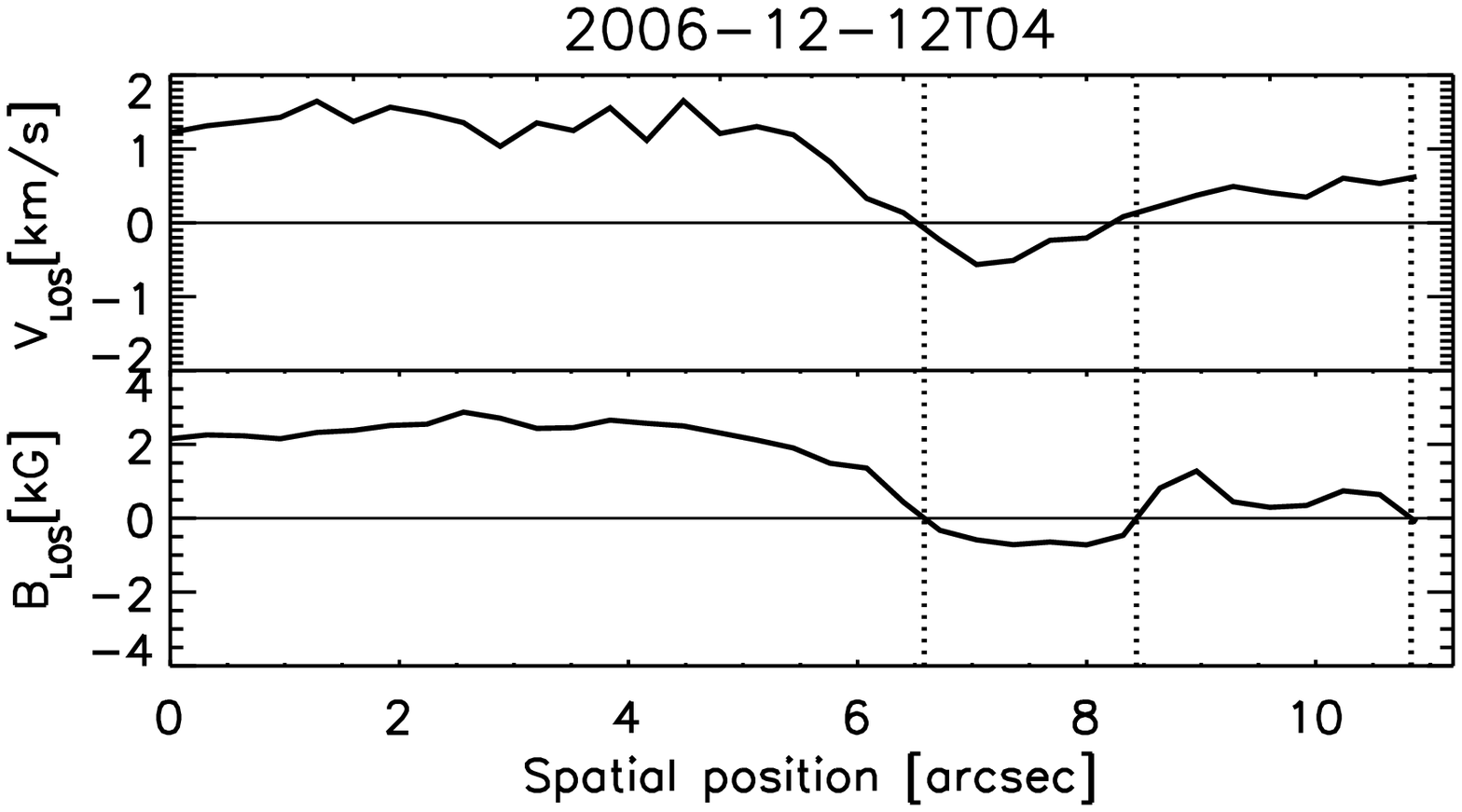}\\
    \includegraphics[width=0.4\textwidth]{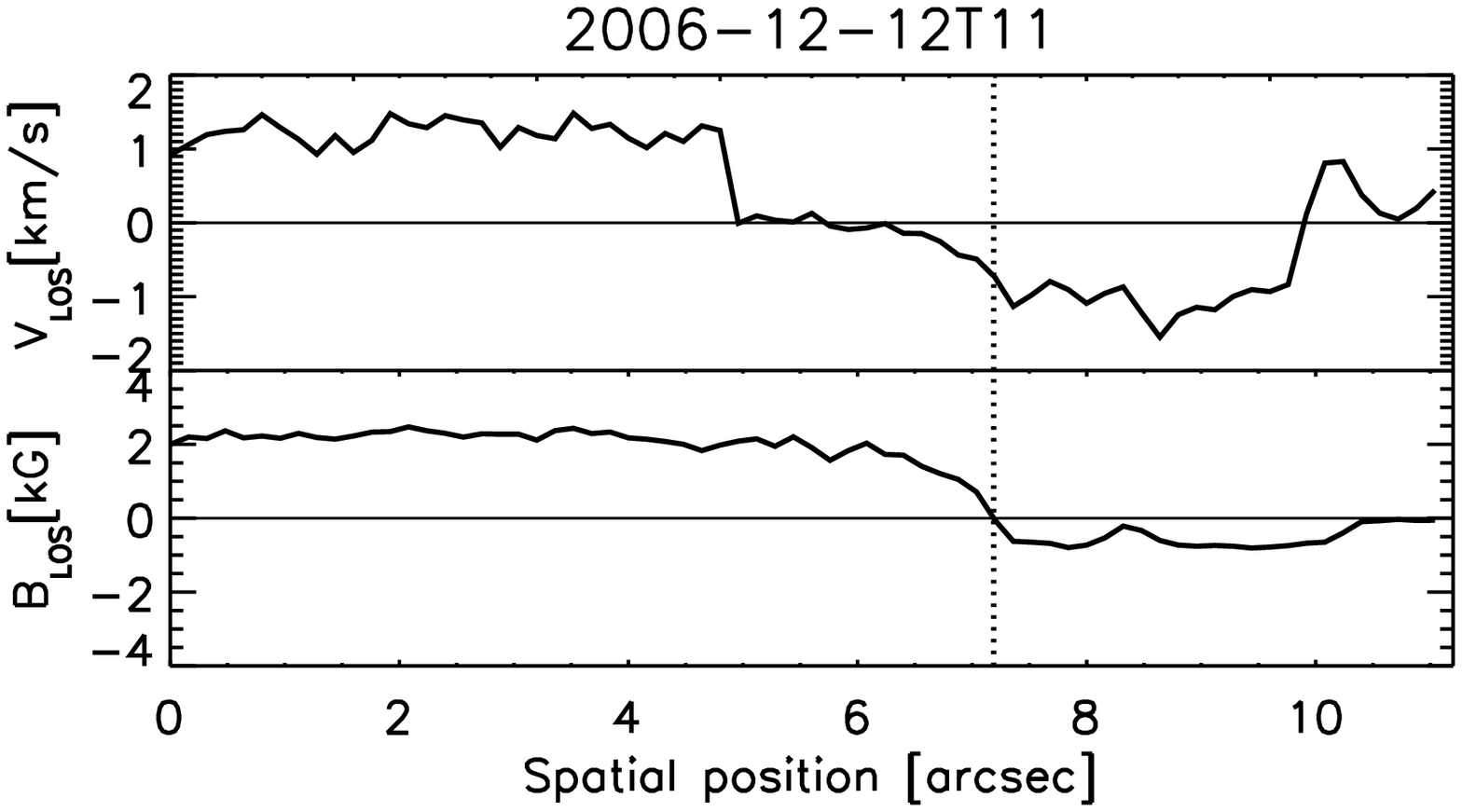}
    \includegraphics[width=0.4\textwidth]{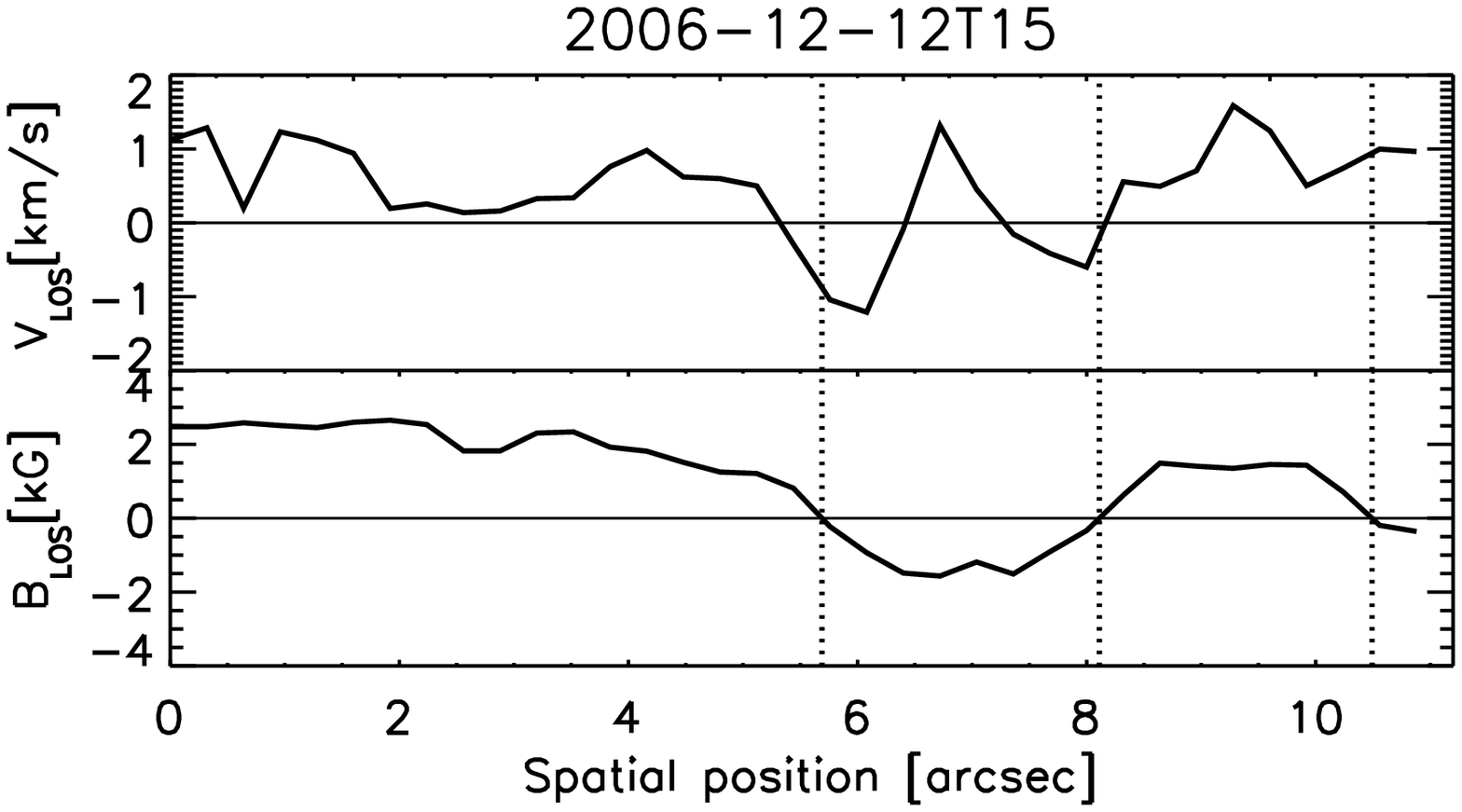}\\
    \caption{{Upper} panel (gray-scaled images) : gray-scaled LOS magnetograms aligned with
    contours of the LOS velocity.
    The blue/orange contours represent upflow/downflow with the contour
    levels of $(-1.0, -0.5, -0.2, \, 0.5, 1.0, 1.5)$ km~s$^{-1}$.
    Black contours indicate the PILs. The FOV of each image is
    $43\arcsec\times16\arcsec$,
    the same as that of the solid rectangle in Figure~\ref{fov_chn}.
    The vertical line in each image indicates the position along
    which the spatial profile of the LOS velocity will be measured.
    {Lower} panel (plots) : spatial profiles of the LOS velocity and the LOS
    magnetic field along the line indicated in each upper panel.
    Dotted lines indicate the polarity reversals.}\label{dopp}
\end{center}
\end{figure}

That the appearance of a flux thread is due to its emergence can be
verified by checking the temporal and spatial variations of Doppler
velocity around the channel structure. Figure~\ref{dopp} shows that
before the appearance of the negative flux thread, at 17 UT on
December 11, most of the area shows downflows slower than \kms{+0.3}.
It is a well-known property that Dopplergrams at active regions
generally show downflows of 0.2--\kms{0.3} \citep{Bha71, How71, How72,
Gio78}. About three hours later at 20UT when the thread forms, upflows
show up at the center of the PIL, just between the positive spot and
the negative thread. This upflow signature is clearer in the spatial
profiles of the LOS velocity across the threads presented in the lower
panel of Figure~\ref{dopp}. At 20 UT on December 11, upflows of about
\kms{-0.3} are seen at the interface between the positive spot and the
negative flux thread. As the time goes on, the signature of upflows
becomes enhanced with the width of the upflow region along the
north--south direction increasing from 0\arcsec.7 (20 UT on December
11) to 3\arcsec.5 (11UT on December 12), and with the upflow speed
increasing from about $-0.3$ to over \kms{-1} in $19$ hr.

Interestingly, the spatial distribution of the LOS velocity around the
negative flux thread at a specific instant, say, at 20 UT on December
11 is consistent with what is expected when an arch-shaped magnetic
flux emerges. We expect that the center of a PIL will display
blueshifts because of the emerging apex and both footpoints on the
opposite sides of the PIL will do redshifts because of the draining
plasma along the field line. In fact we see faster downflows of
\kms{+1.5} at both ends of the negative flux thread and slower upflows
of \kms{-0.3} at the center of the PIL.

\begin{figure}[tb]
\begin{center}
    \includegraphics[width=0.45\textwidth]{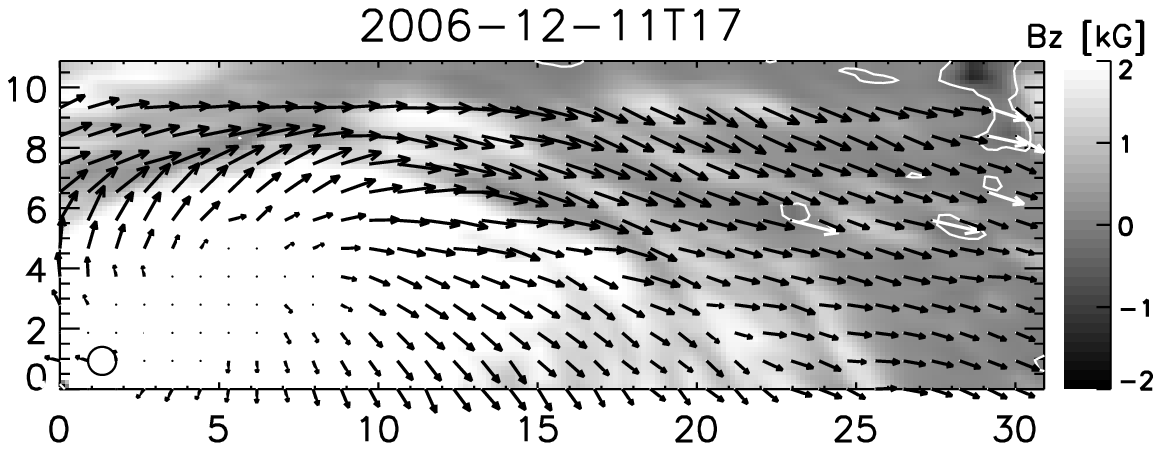}
    \includegraphics[width=0.45\textwidth]{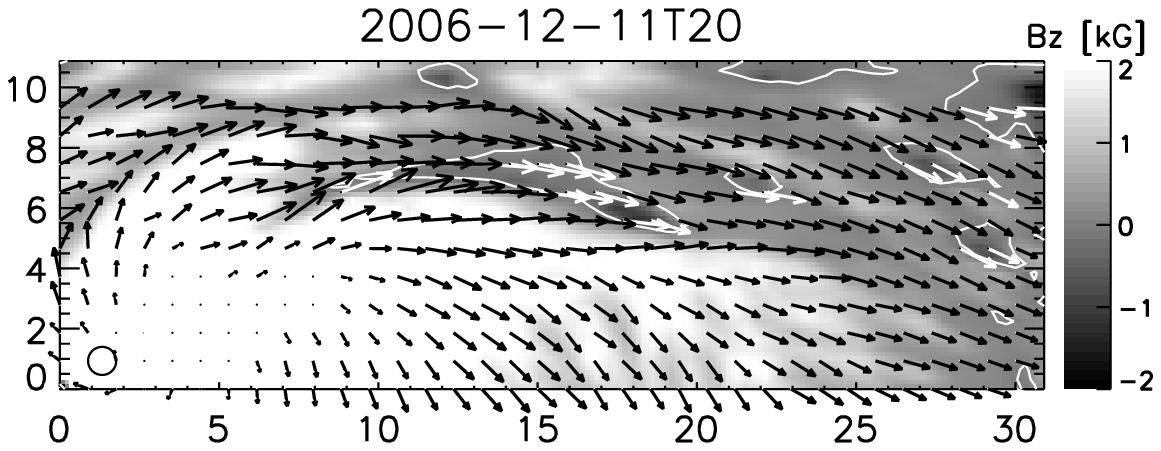}\\
    \includegraphics[width=0.45\textwidth]{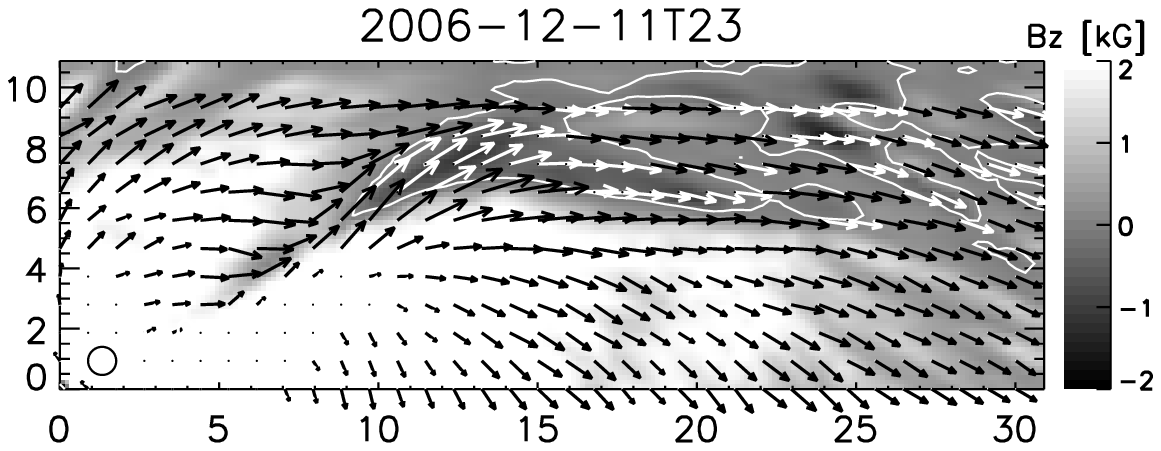}
    \includegraphics[width=0.45\textwidth]{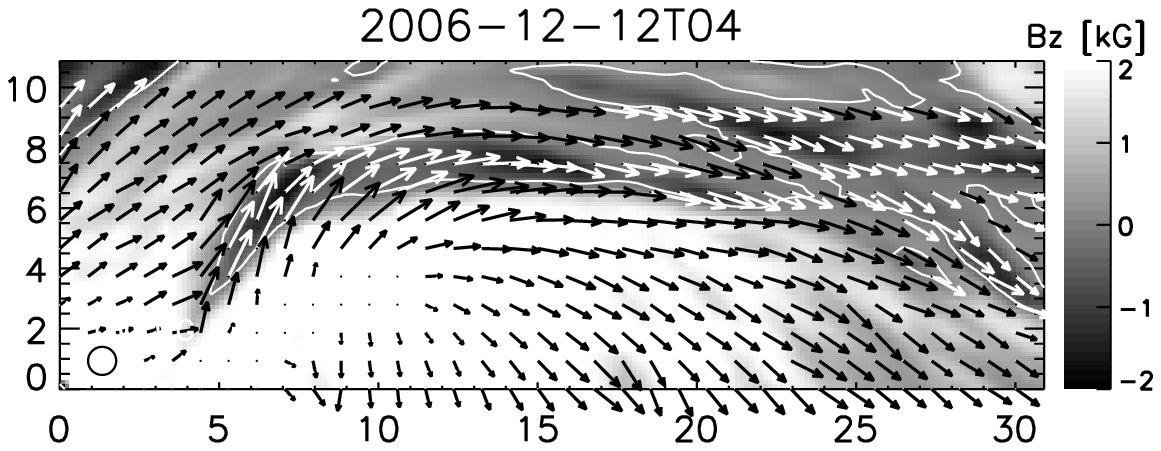}\\
    \includegraphics[width=0.45\textwidth]{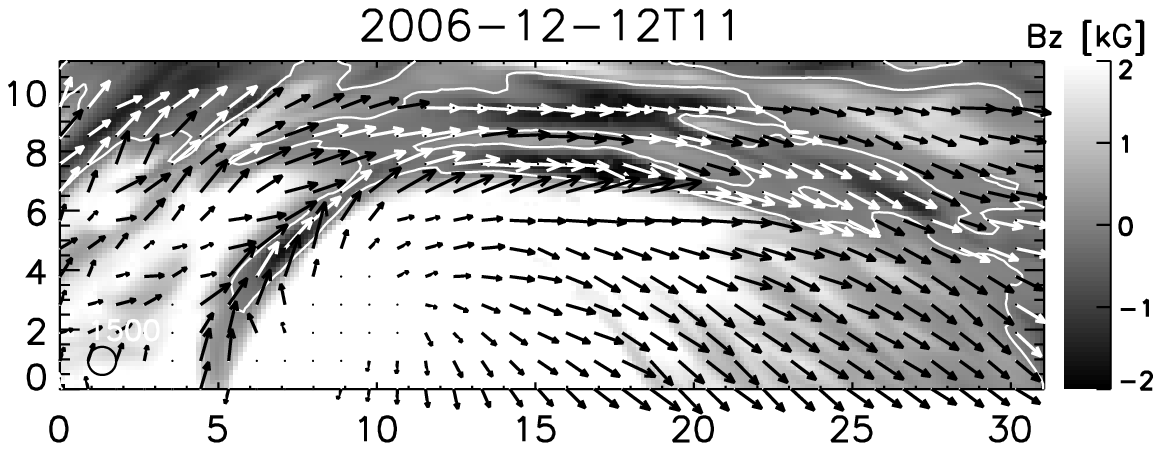}
    \includegraphics[width=0.45\textwidth]{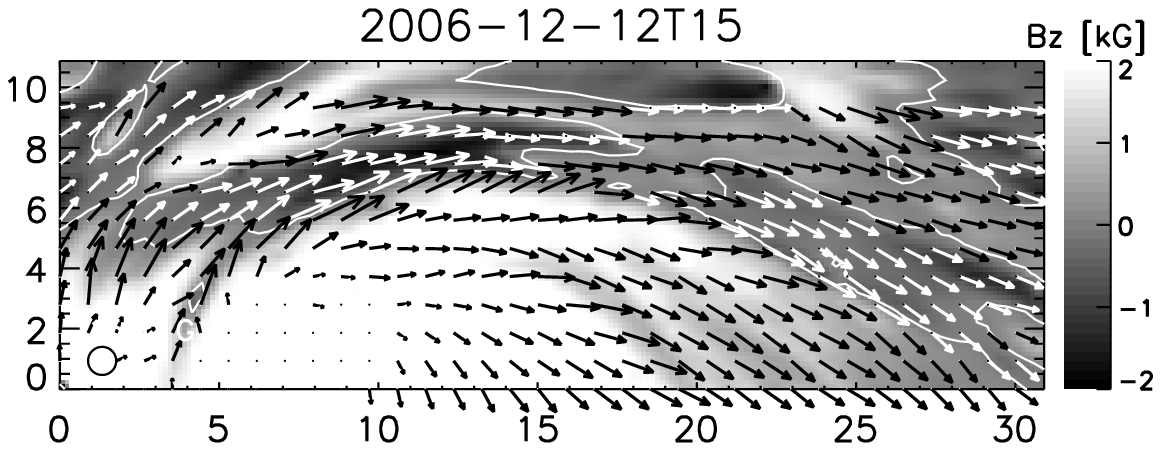}\\
    \caption{Transverse magnetic field vectors superposed on gray-scaled LOS magnetograms.
    The unit length of the arrow (the diameter of the circle at the lower left corner)
    corresponds to the transverse field strength of $2000$~G.
    The color of arrows was set to be opposite to that of the background gray-scaled magnetogram.}
    \label{vectogram}
\end{center}
\end{figure}

With the result of the ME inversion of SP data, we have checked the
transverse vector magnetic field and inclination angles around the
magnetic channel to figure out the magnetic field configuration of the
negative flux thread. Vector magnetograms presented in
Figure~\ref{vectogram} clearly show that the magnetic field around the
channel structure is predominantly horizontal and nearly parallel to
the channel. Horizontal fields are as strong as 3000 G while vertical
fields are weaker than 1000 G. The shear angles and inclinations of
this channel area at a specific time were previously studied by
\citet{wang08}. Here we focus on the beginning phase of the flux
emergence and thereafter. The transverse magnetic field vectors at 17
UT on December 11 diverge out of the positive spot, and tightly wind
around the umbra clockwise, indicating that this active region is
already highly non-potential. Before the channel formation the azimuth
of the vectors is smooth and nearly uniform, but as the negative flux
thread emerges and grows, it steeply varies across the thread.
Combined with the change of magnetic polarity, this spatial variation
of azimuth is consistent with a magnetic configuration of negative
(left-handed) twist. Considering that the transverse field is already
nearly parallel to the channel at the early phase of the flux
emergence and that persistent upflows exist along the thread, we think
that the twist is continuously carried by the emerging flux from below
the surface rather than being induced by the photospheric flows after
emergence.

\begin{figure}[tb]
\begin{center}
    \includegraphics[width=0.45\textwidth]{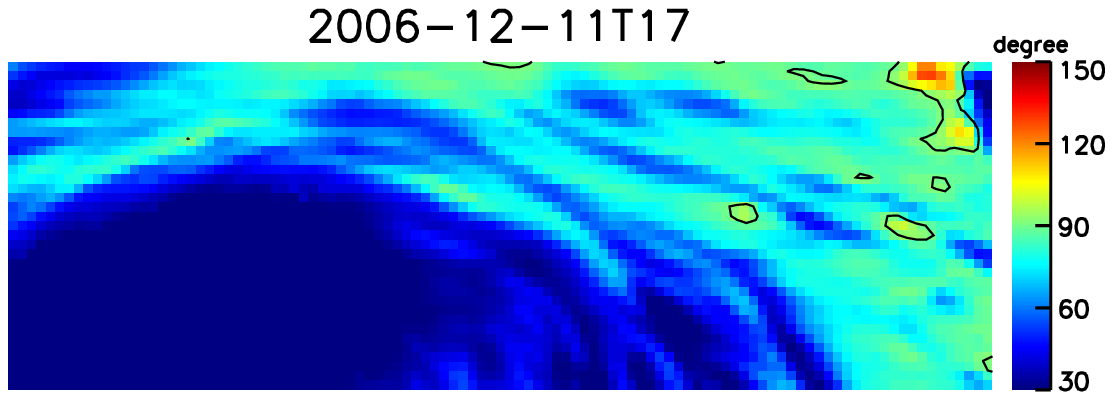}
    \includegraphics[width=0.45\textwidth]{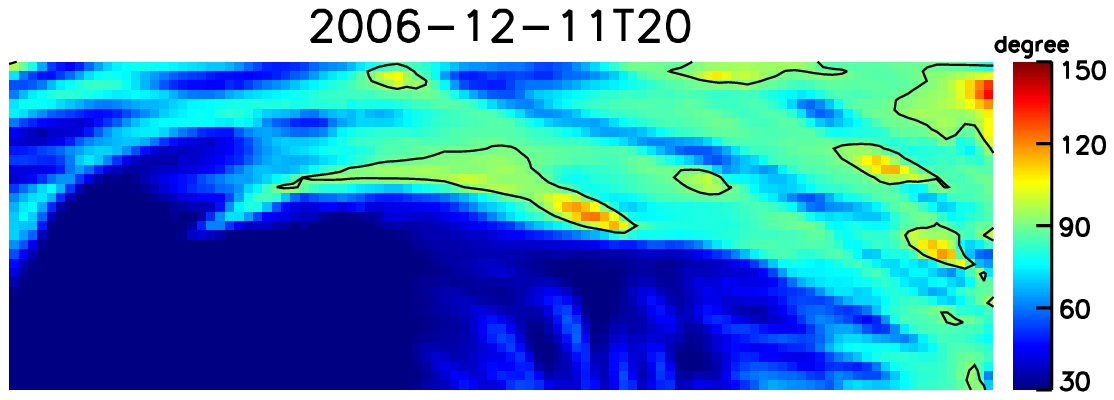}\\
    \includegraphics[width=0.45\textwidth]{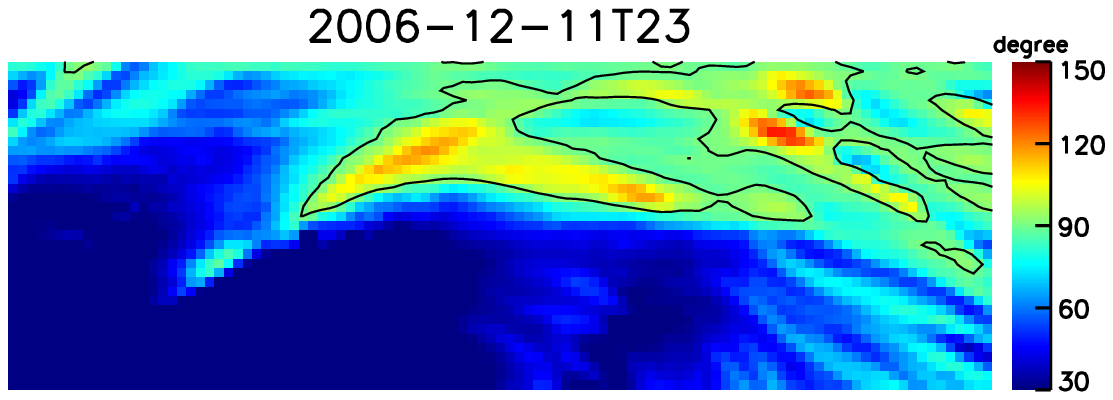}
    \includegraphics[width=0.45\textwidth]{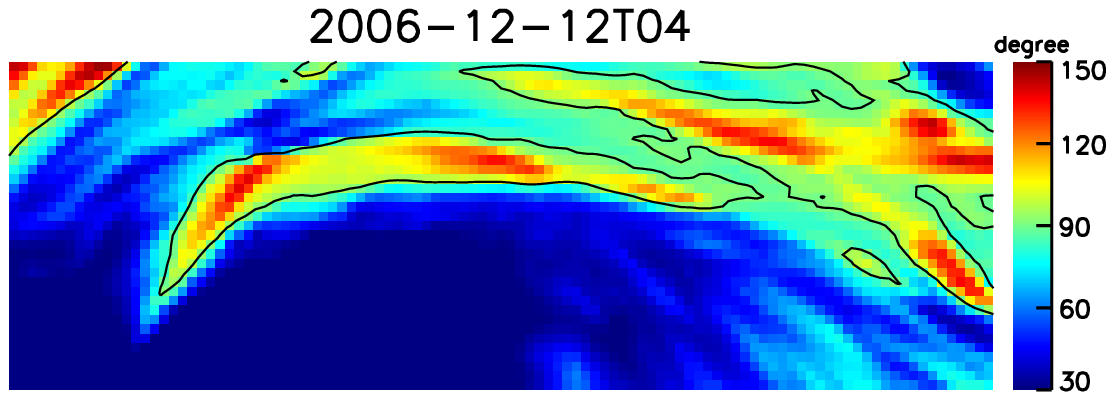}\\
    \includegraphics[width=0.6\textwidth]{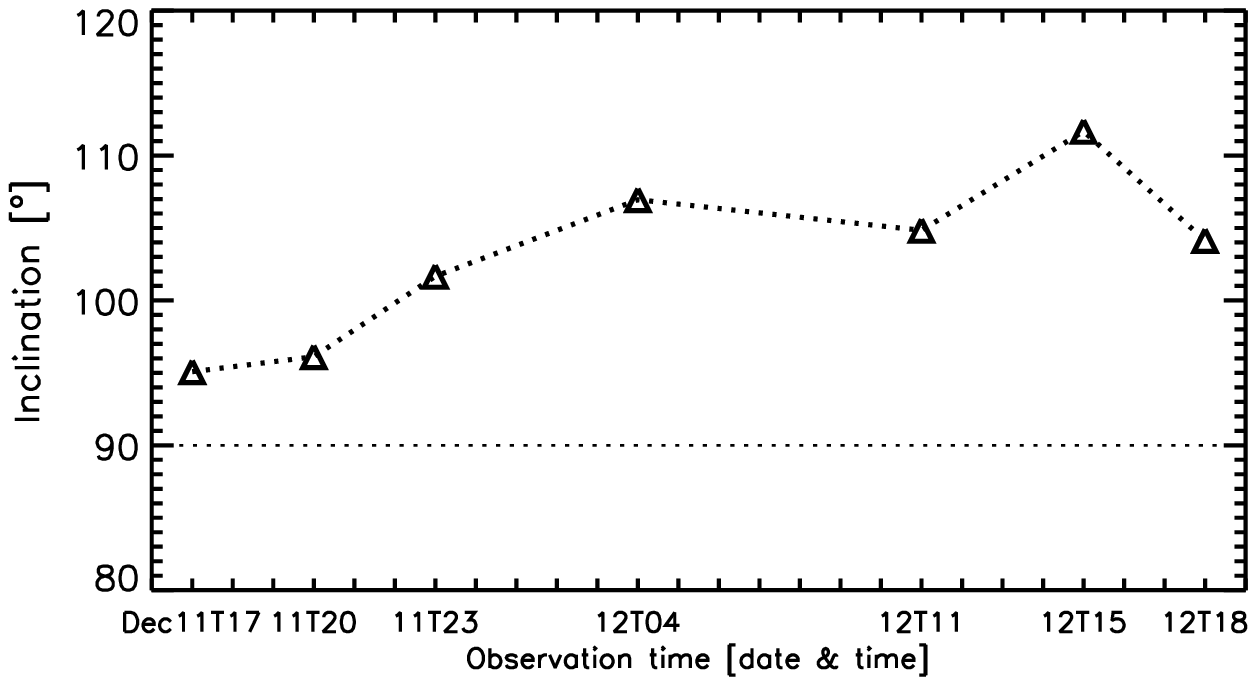}\\
    \caption{{Upper} panel : spatial distribution of the inclination angle.
    An angle of $90^{\circ}$ corresponds to horizontal fields
    to the surface. Angles either less than $30^{\circ}$ or over
    $150^{\circ}$ were suppressed in the map in order to describe
    the change in inclination around the channel more clearly.
    The inclination angles of the channel are all within
    $30^{\circ}--150^{\circ}$ range.
    The black solid contours indicate the PIL.
    {Lower} panel : time variation of the inclination angle that was measured and
    averaged along each negative flux thread.}\label{incli}
\end{center}
\end{figure}

Figure~\ref{incli} shows that the direction of magnetic field changes
from horizontal to vertical at the channel between the positive spot
and the negative thread, as the negative flux thread grows. At 17 UT
on December 11, before the formation of the magnetic channel most of
the region shows nearly $90^{\circ}$ inclination indicating that the
magnetic field is almost parallel to the surface. After the thread
emerges, the inclination gradually increases and at some places it
reaches about $120^{\circ}$ already at 04 UT on December 12. This
trend of increase is clearly shown in the time profile of inclination
spatially averaged along the negative flux thread. Roughly speaking,
the averaged inclination  increases from $95^{\circ}$ to $112^{\circ}$
for 22 hr, which means that the magnetic field gradually changes to
vertical to the surface. The inclination angle of $112^{\circ}$ means
that the field is {negative} and $22^{\circ}$ inclined from the
surface. Together with results on the Doppler velocity described
above, this result supports the emergence of arch-shaped magnetic
fields.

\begin{figure}[tb]
\begin{center}
    \includegraphics[width=0.45\textwidth]{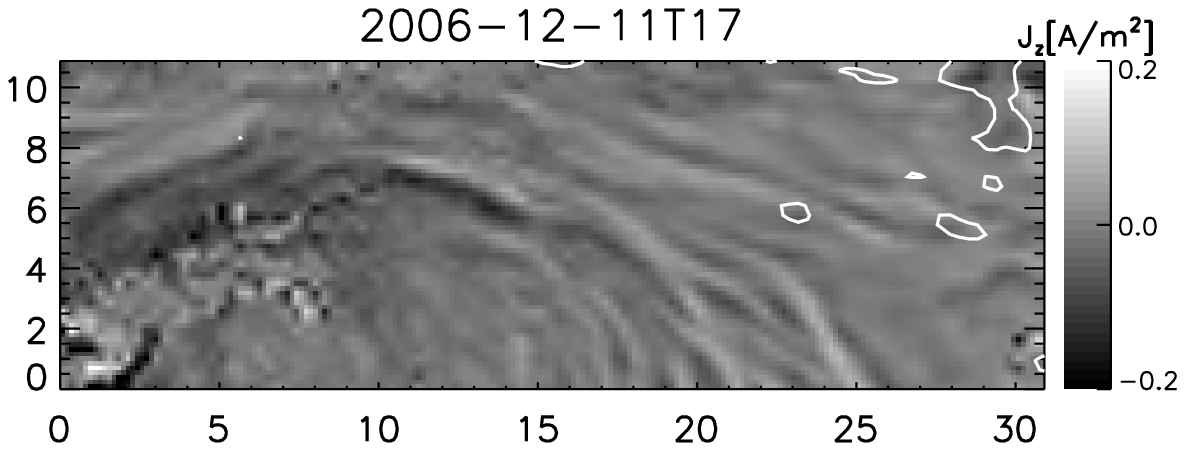}
    \includegraphics[width=0.45\textwidth]{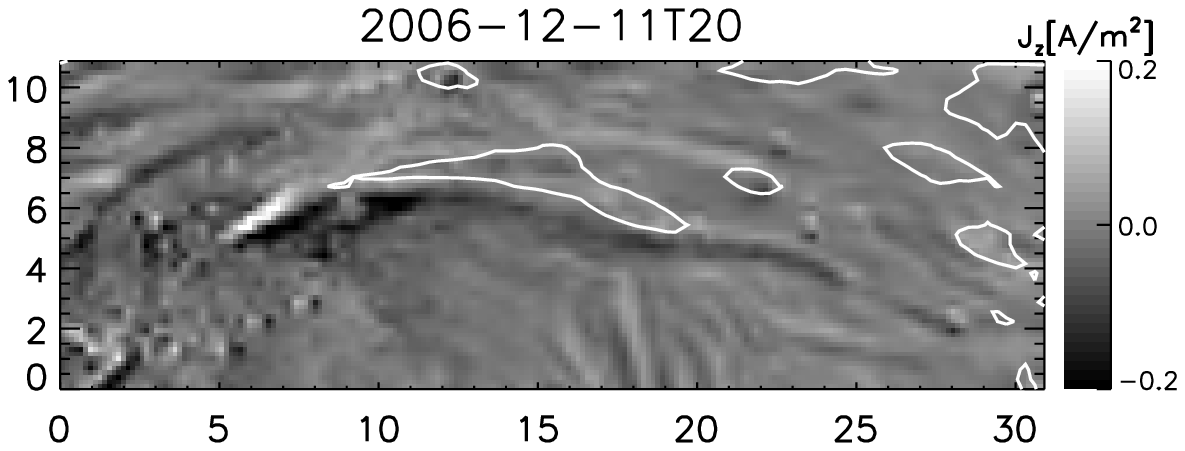}\\
    \includegraphics[width=0.45\textwidth]{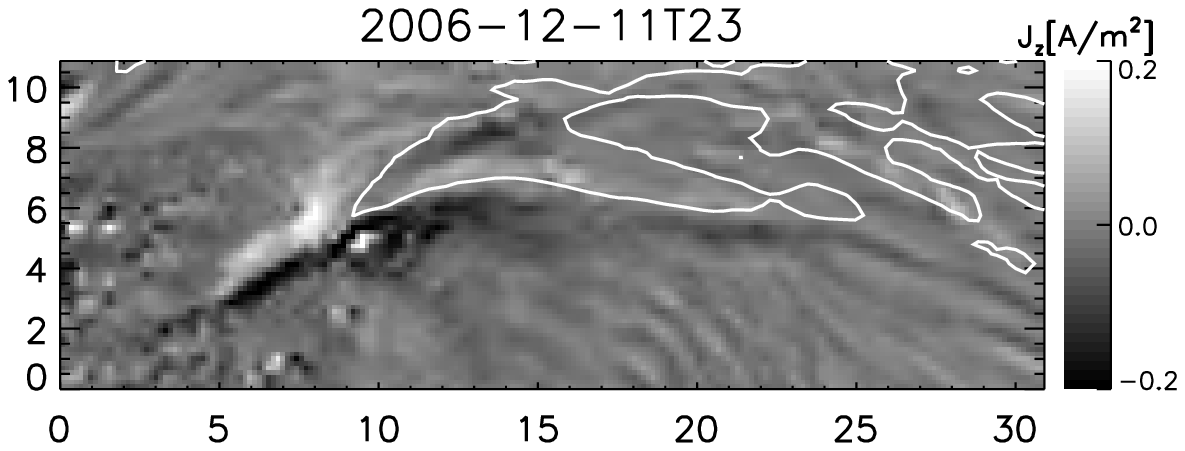}
    \includegraphics[width=0.45\textwidth]{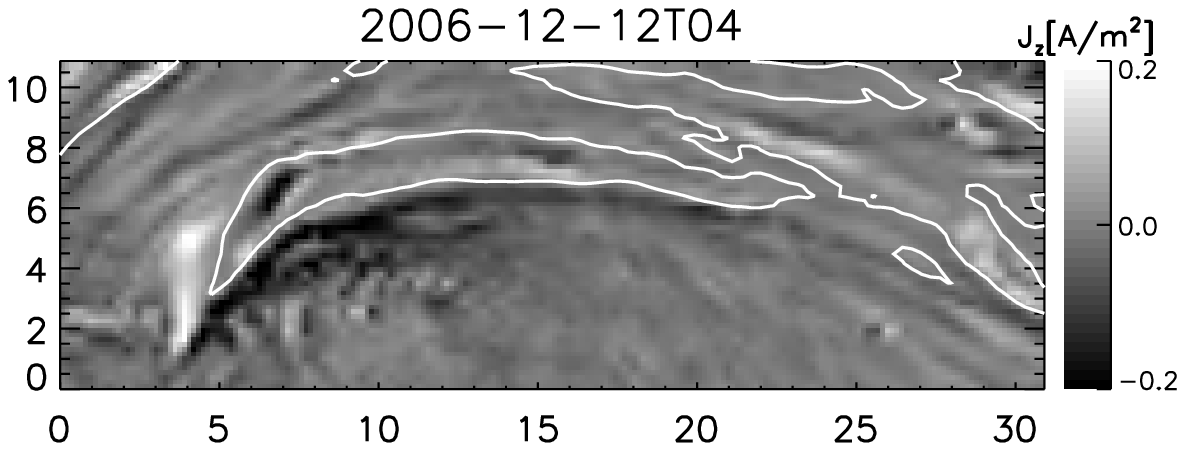}\\
    \includegraphics[width=0.45\textwidth]{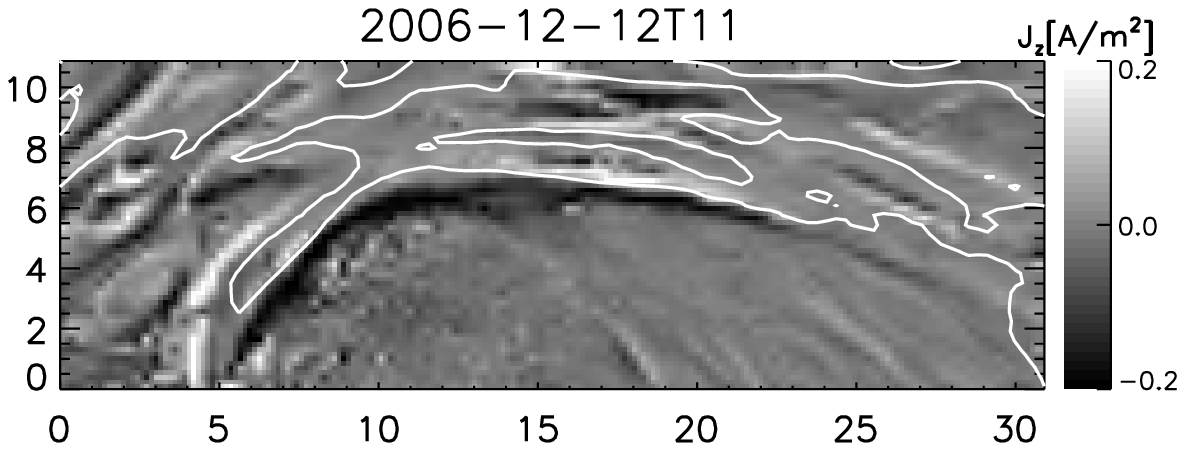}
    \includegraphics[width=0.45\textwidth]{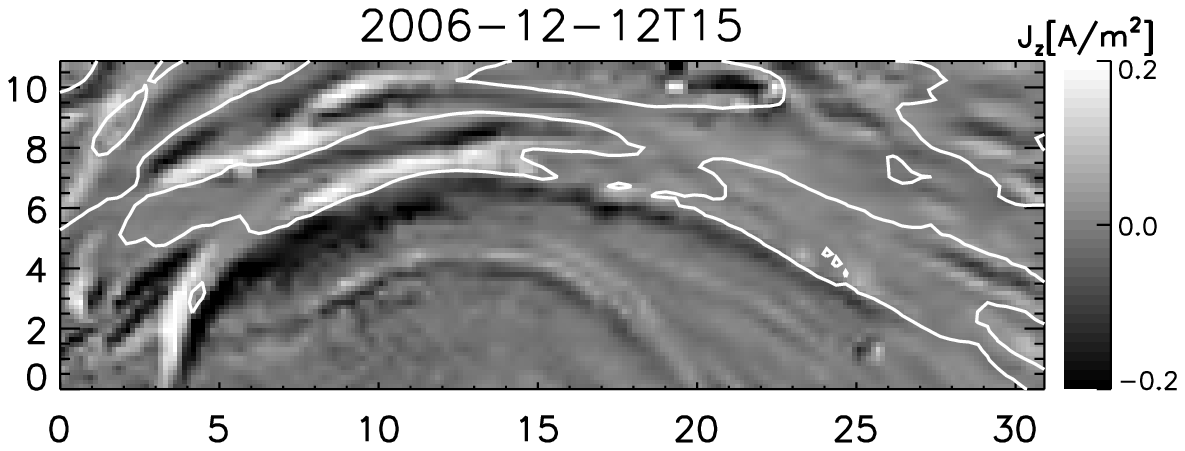}\\
    \caption{Maps of the vertical components of electric current (gray scale) with the
    PILs (white contour). The positive sign
    corresponds to the upward direction and the negative to the
    downward. The current field stronger than $\pm 0.2$~A~m$^{-2}$ was
    suppressed in the image display.}
     \label{current}
\end{center}
\end{figure}

Figure~\ref{current} shows that the vertical component of electric
current density is strong along the PIL between the negative flux
thread and the positive penumbral magnetic field. The negative flux
thread is identified by the enclosing white contour in each panel. The
most prominent is the appearance and growth of a pair of current
threads of opposite polarity: a negative one in the magnetic region of
positive polarity and a positive one in the region of the negative
flux thread. Its signature is seen as early as 17 UT on December 11,
even before the emergence of the negative flux thread. This pair of
current threads is another manifestation of the steep gradient of the
transverse component of magnetic field across the flux thread, and is
quite consistent with the emergence of a horizontal flux tube that is
twisted in the left-handed sense. A careful examination of
Figure~\ref{current} reveals that after the appearance of the negative
flux thread, the current thread of negative polarity gets stronger
than the one of positive polarity. This asymmetry may be explained by
the interaction of the newly emerging field and pre-existing overlying
field. Since the current thread of negative polarity shows up in the
region of stronger field, the magnetic flux should be more squeezed,
resulting in a steeper magnetic gradient and a higher electric current
density.

{An opposite pattern of current density is also found in
Figure~\ref{current}: a negative current in the magnetic region of
negative polarity, and a positive one in the negative magnetic region.
It could be interpreted in two ways. The weaker current threads with
the opposite direction to one along the magnetic channel may represent
the return current of the twisted flux tube. In the case of the
typical twisted flux tube, the surface current flows opposite to the
field-aligned volume current so that the twisted flux tube could have
a finite size. They are concentrated in the vicinity of the flux
thread at the early phase of the flux emergence. The other possible
interpretation is that those opposite current threads may indicate
that local flux is twisted in the right-handed sense, which is
opposite to the twist of magnetic channel structure. It is not
surprising if some local magnetic fields emerge carrying opposite
helicity to the total helicity of the active region \citep{Kus04,
Cha10}. However, the current threads are fragmented and get complex as
the magnetic channel evolves and more fine-scale flux threads emerge
nearby. Therefore, it is not easy to interpret current distribution at
the later phase of the channel formation.}

\subsection{Coronal Magnetic Field}
\begin{figure}[tb]
\begin{center}
    \includegraphics[width=0.35\textwidth]{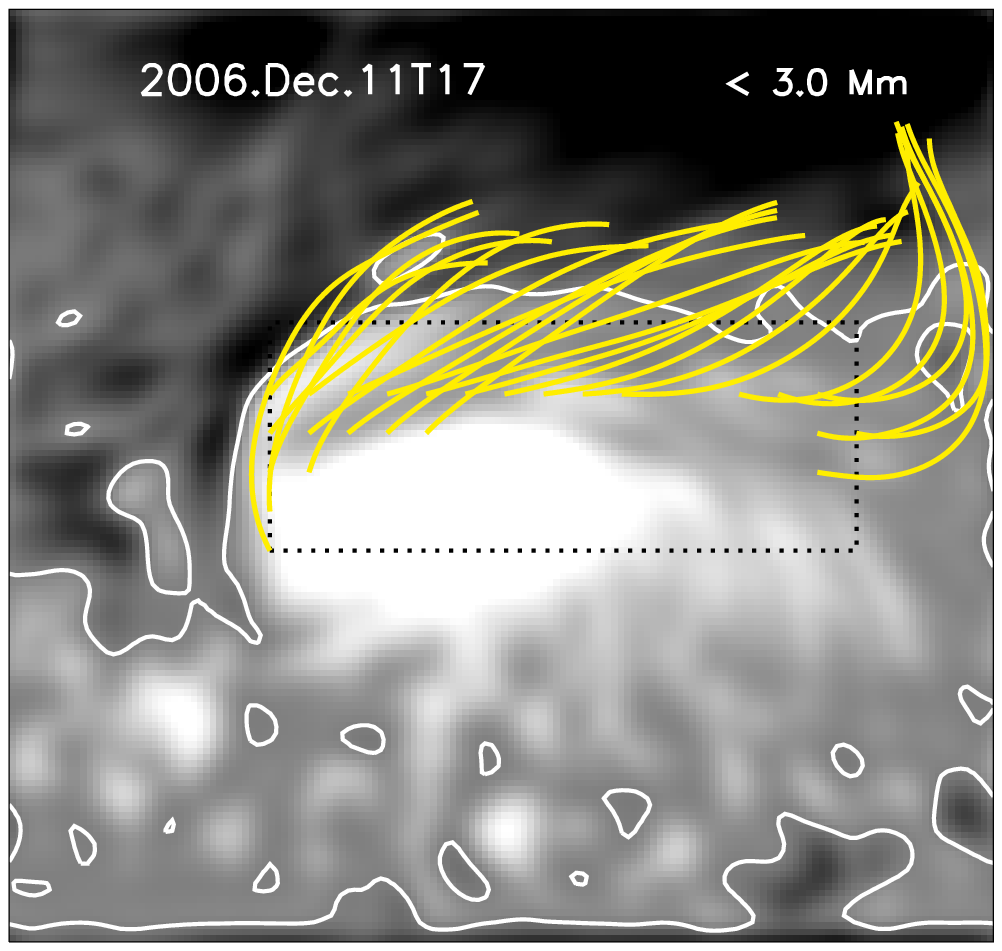}
    \includegraphics[width=0.45\textwidth]{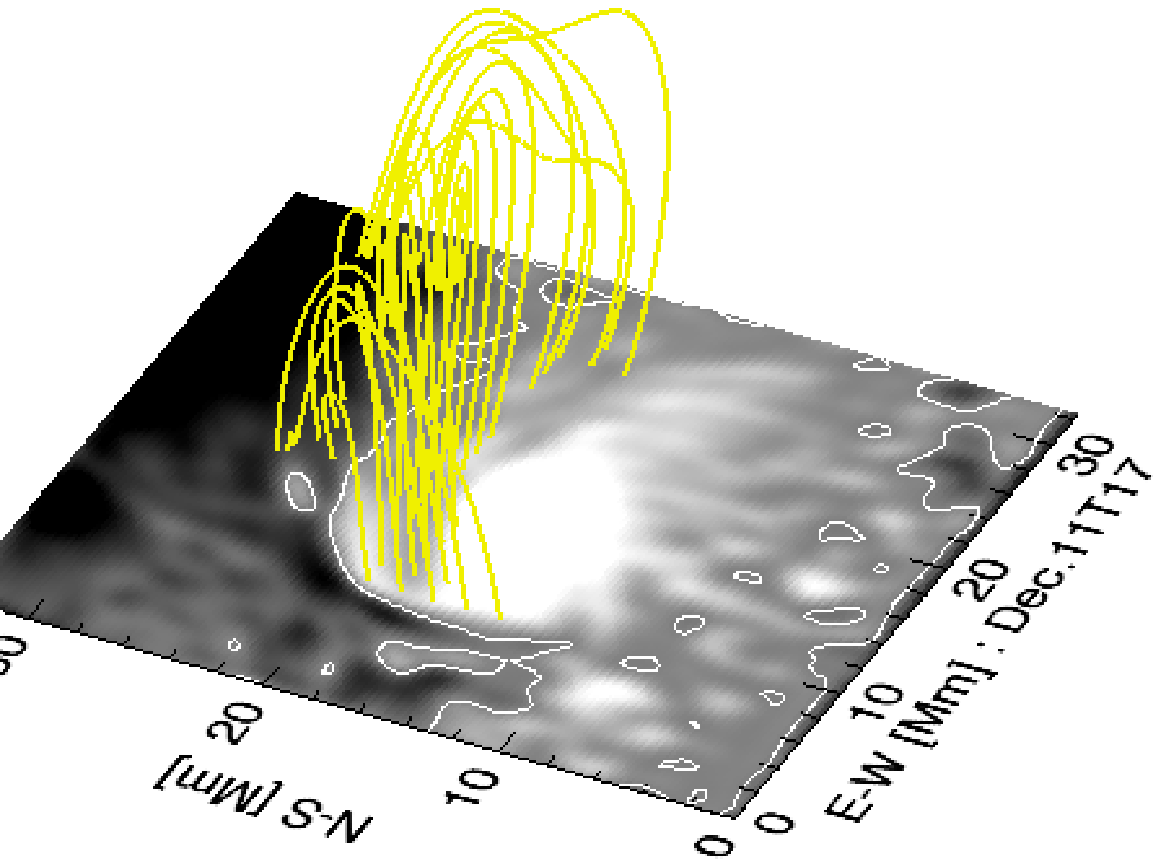}\\
    \includegraphics[width=0.35\textwidth]{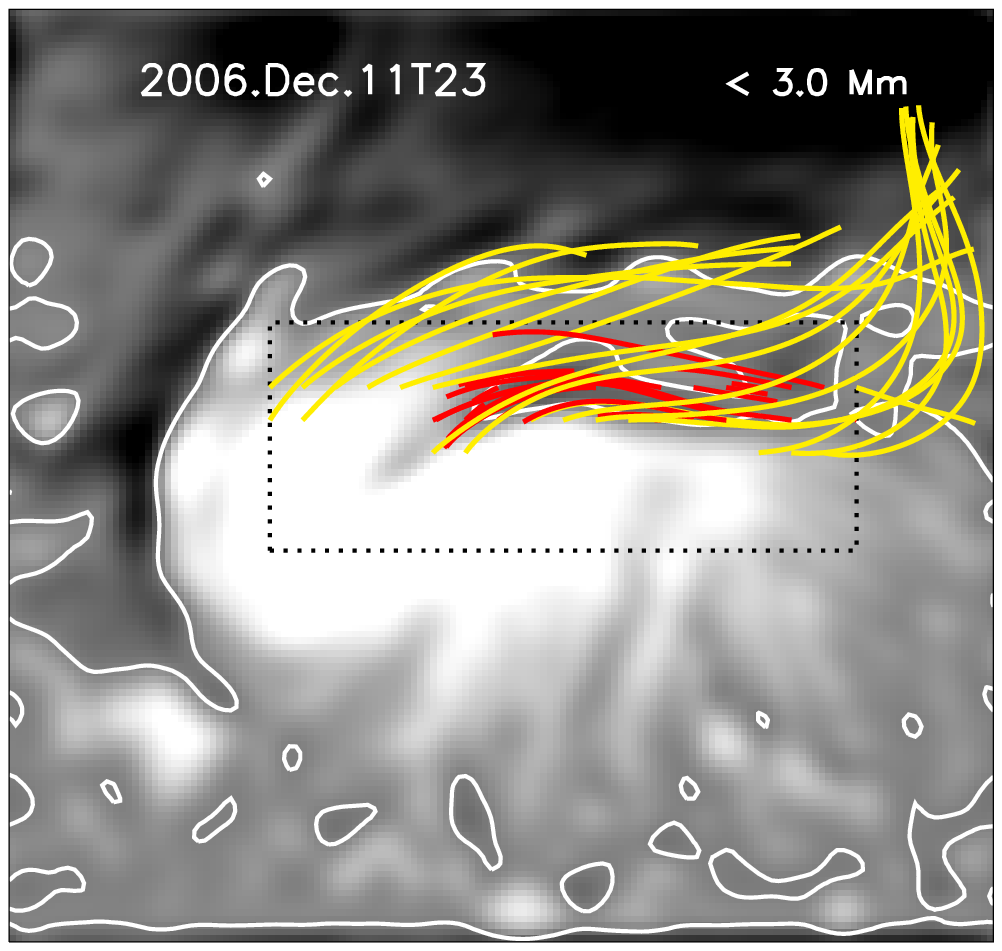}
    \includegraphics[width=0.45\textwidth]{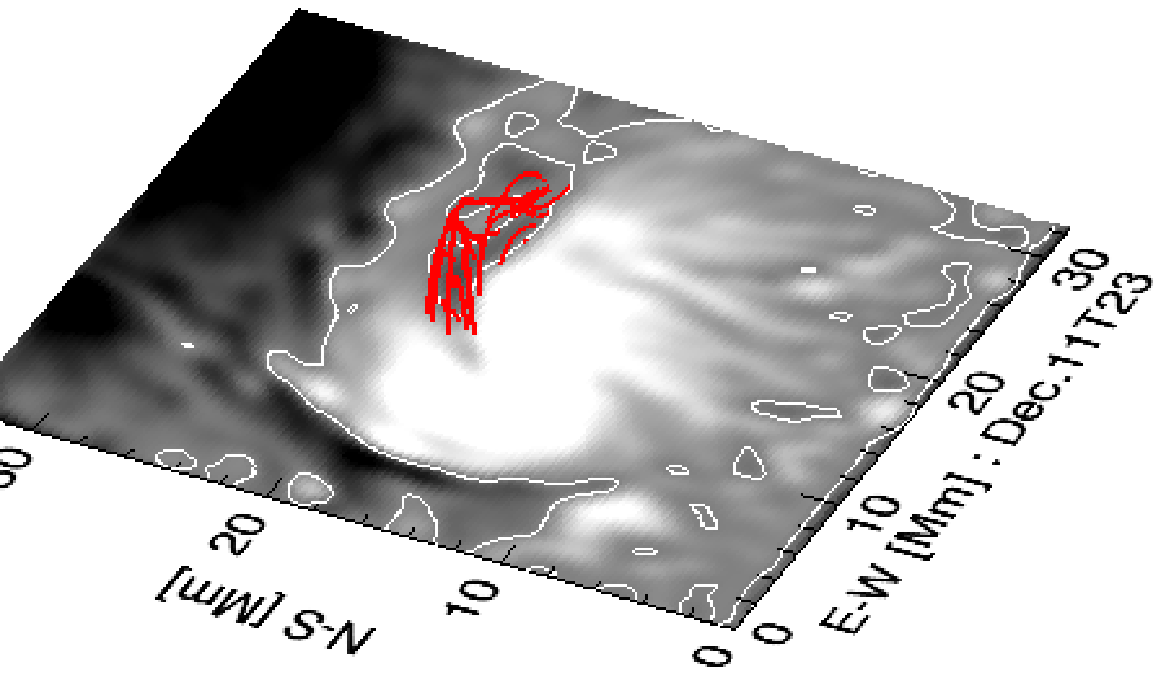}\\
    \includegraphics[width=0.35\textwidth]{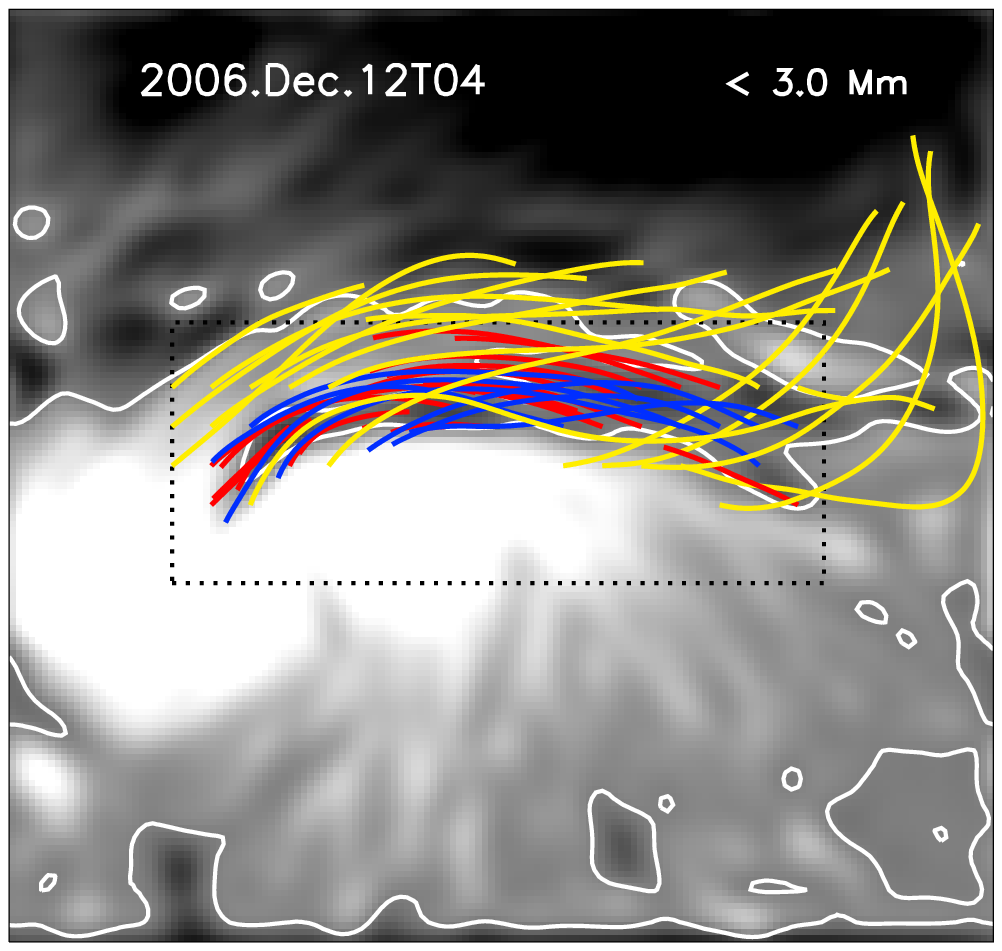}
    \includegraphics[width=0.45\textwidth]{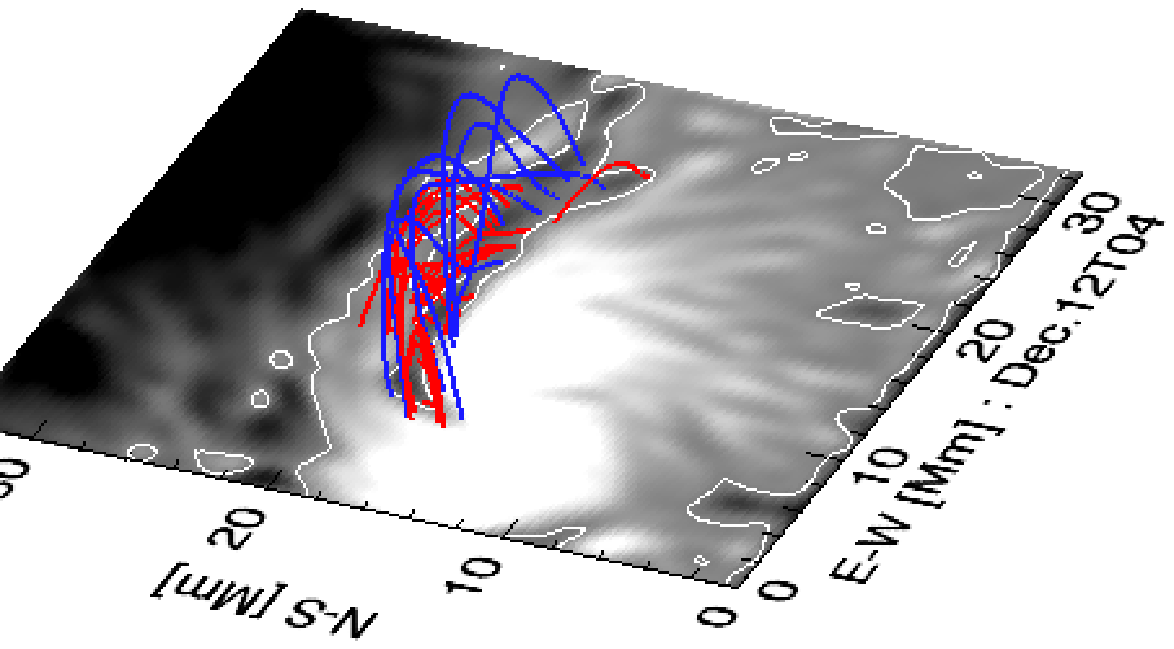}\\
    \caption{Magnetic field lines from the NLFFF extrapolation
    viewed from the top ({left} panel) and side ({right} panel).
    Field lines lower than 3~Mm only are displayed co-aligned with the vertical
    magnetic field (gray scale images). Field lines of the channel structure
    are colored either red or blue depending on their height, and
    overlying field lines are colored yellow. Red indicates field
    lines lower than 1~Mm, blue higher than 1~Mm and lower than
    2~Mm. Overlying field lines are omitted in side view
    images to show the channel field lines clearly.} \label{3Dlines}
\end{center}
\end{figure}

With the aid of the NLFFF extrapolation, we have checked the temporal
change of the coronal magnetic field and the three-dimensional current
field above the magnetic channel, assuming that the evolution of the
magnetic field in the corona may be approximated to be quasi-static
and force-free. Figure~\ref{3Dlines} displays the field lines that are
lower than 3~Mm and near the negative flux thread at each instant. The
figure clearly demonstrates the emergence of highly sheared field
lines (red lines and blue lines) along the photospheric negative flux
thread inside a pre-existing less sheared arcade (yellow lines).
Before the appearance of the negative flux thread, at 17 UT on
December 11, most of the field lines are over 3~Mm (omitted from the
figure) and the lower field lines are connecting the positive and the
negative spots. After the negative flux thread emerged, at 23UT, new
field lines lower than 1~Mm connecting the positive magnetic region
and the negative flux thread (red lines) are observed below yellow
lines at 23 UT on December 11. At 04 UT on December 12, the flux
thread has grown laterally and blue lines slightly higher than red
lines are visible along the flux thread. Since we did not trace the
field line footpoints along the time, we could not tell which field
line corresponds to which among different observation times. However,
since the negative flux thread is continuously emerging, it seems
natural to assume that some of the red lines at 23 UT on December 11
extended upward and are seen as blue lines at 04 UT on December 12.
The overall topology of field lines at 04 UT on December 12, more
sheared field lines inside less sheared ones, resembles the upper
portion of a twisted flux tube that has partly emerged through the
photosphere. These results from NLFFF extrapolations are consistent
with the picture of the emergence of a twisted flux tube. Although
\citet{Sch08} and \citet{Mag08} also showed evidence of the emerging
twisted flux tube in AR 10930, we are looking at somewhat different
region in that the scale size of the channel structure is a lot
smaller than theirs.

\begin{figure}
\begin{center}
    \plottwo{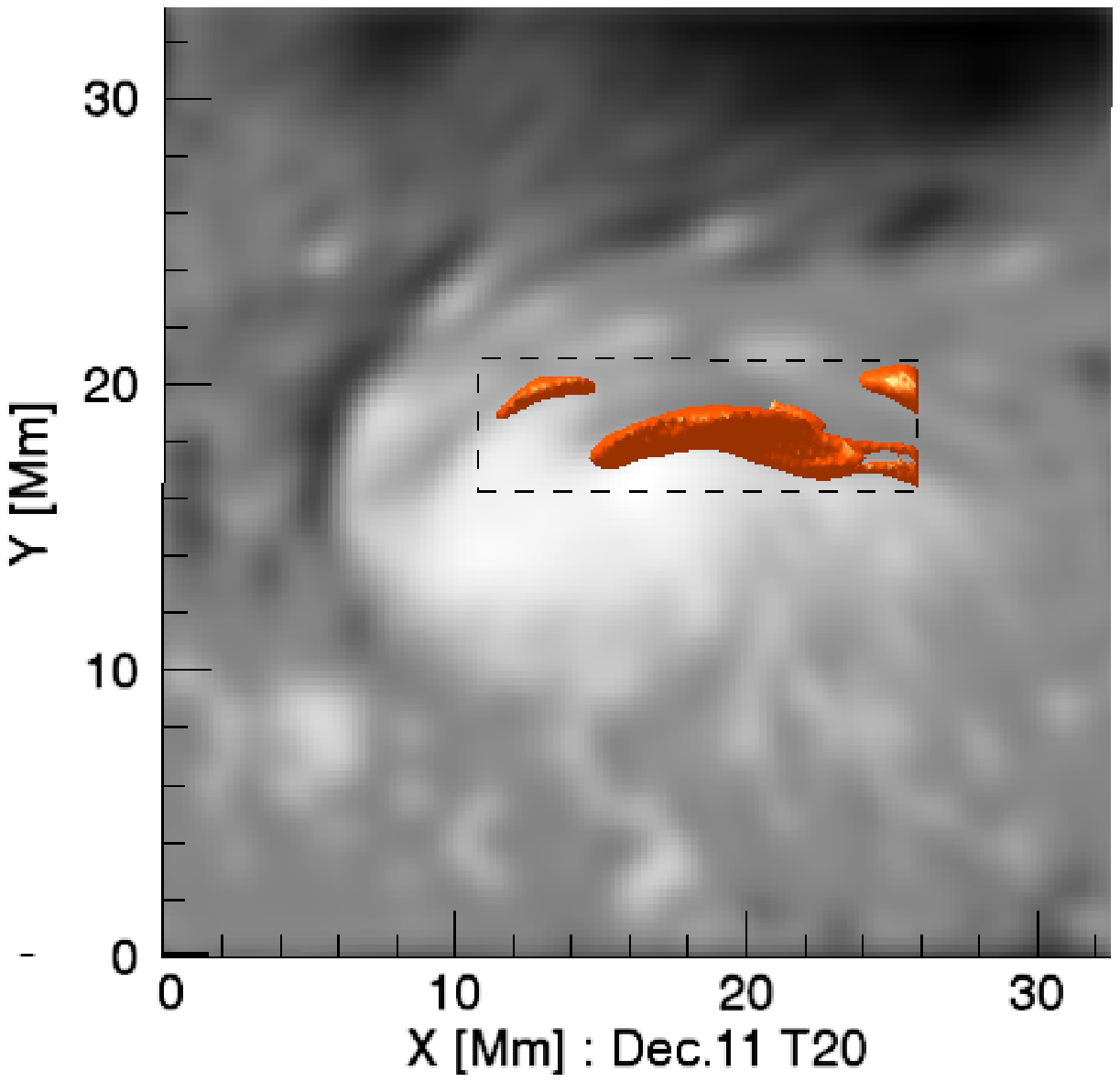}{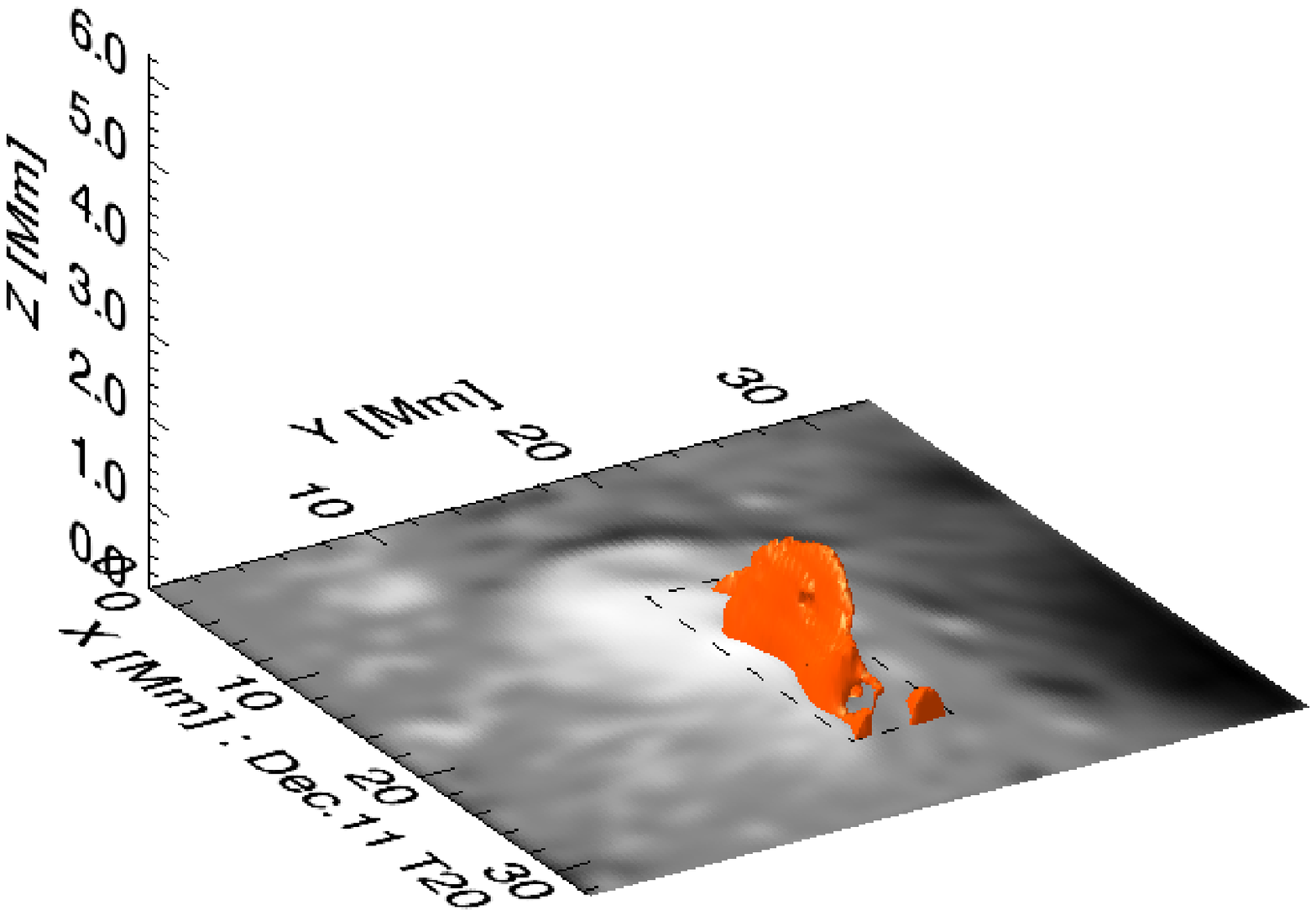}\\
    \plottwo{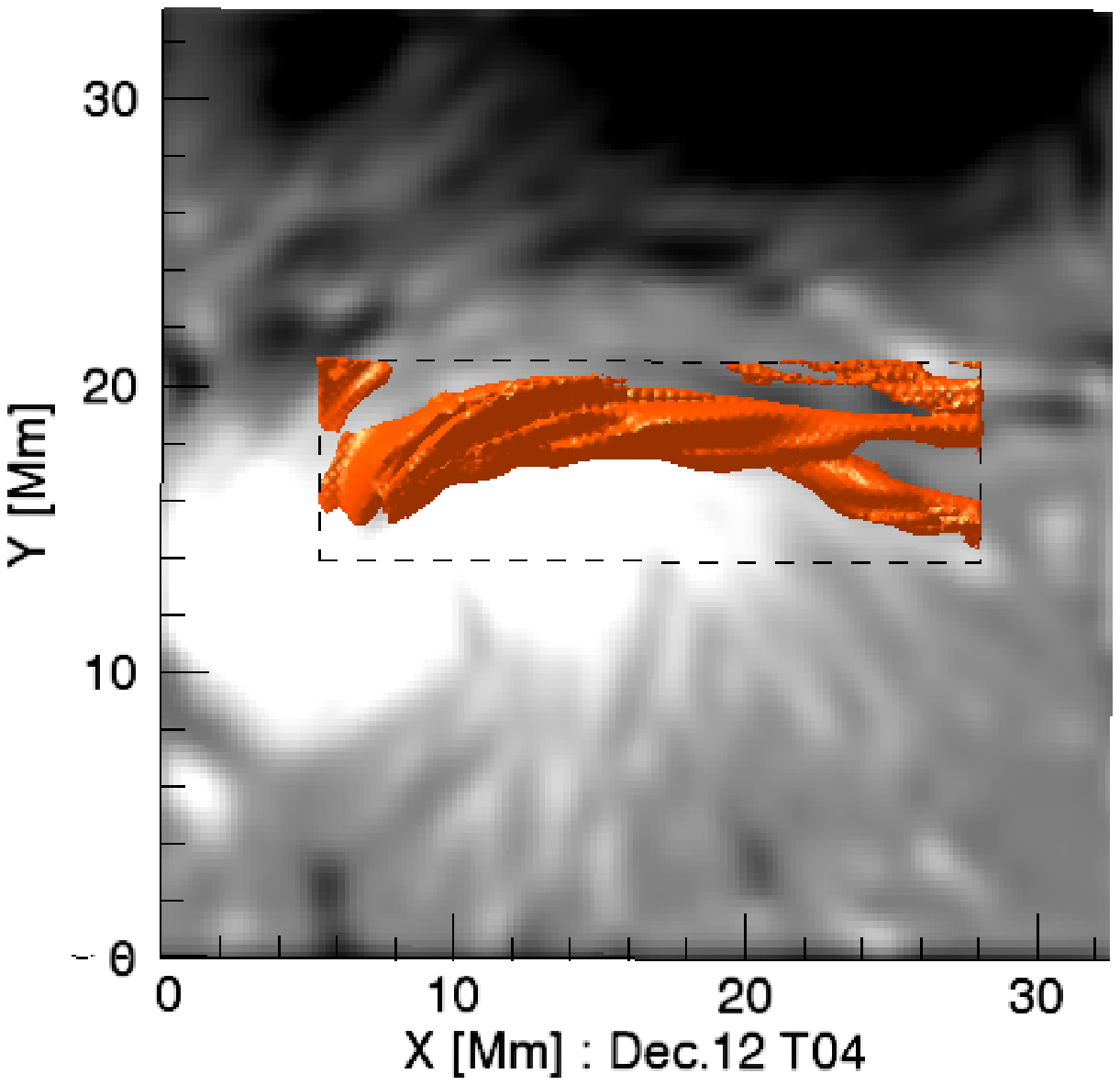}{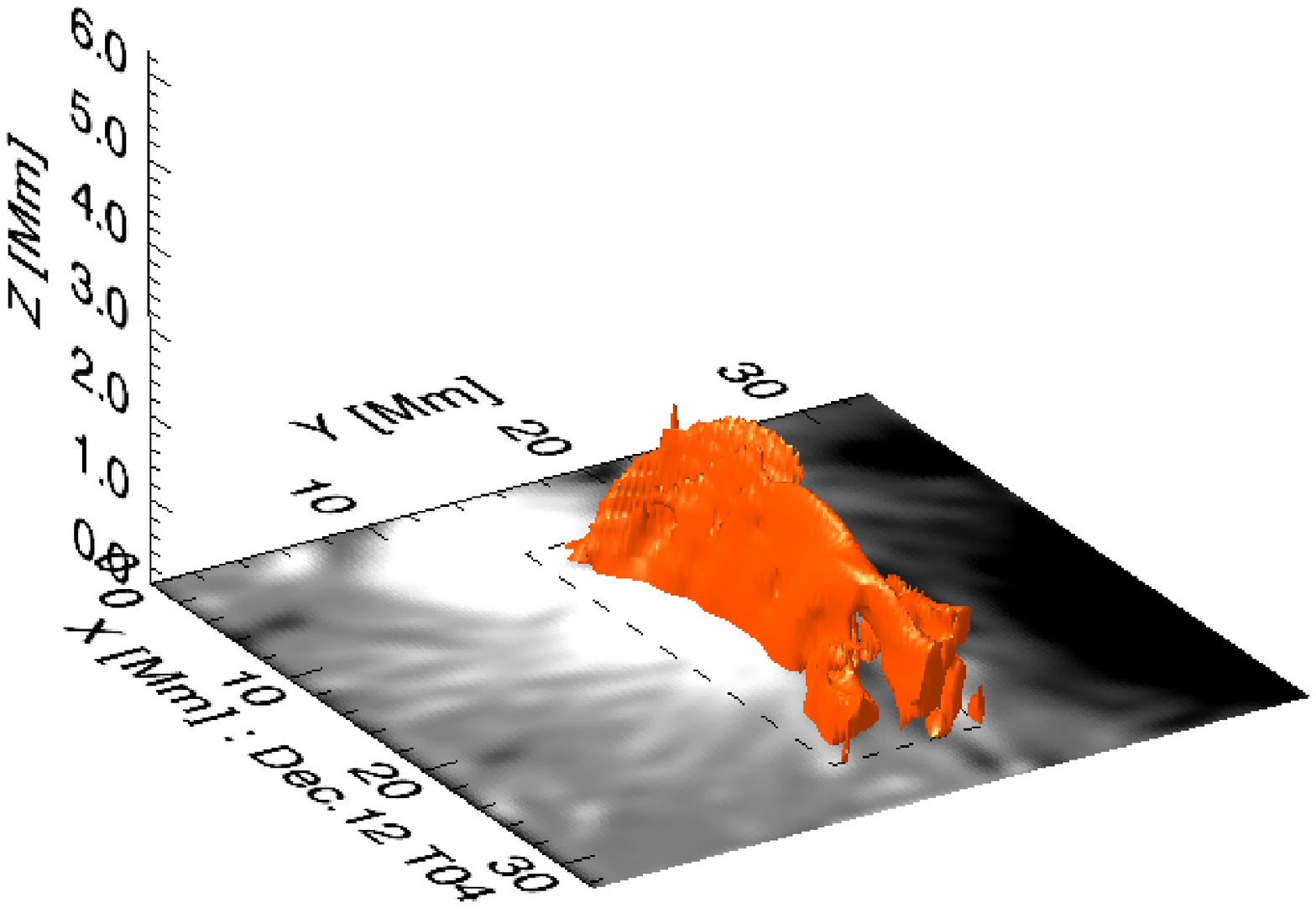}\\
    \caption{Isosurfaces of the transverse component of current density,
    $|J_{t}|=\sqrt{{J_{x}}^{2}+ {J_{y}}^{2}}=0.1~\textrm{A}~\textrm{m}^{-2}$,
    viewed from the top (left) and side (right).
    The upper panel was taken at 20 UT on December 11 and the lower panel
    at 04 UT on December 12.
    Current density out of the rectangle presented by the dashed line
    was suppressed in the display to show the
    current density along the channel only.}
    \label{3Dcur}
\end{center}
\end{figure}

The NLFFF extrapolations allow us to calculate all the components of
electric current density in the lower corona. Figure~\ref{3Dcur} shows
the isosurfaces of the transverse component of current density with
the value of \amm{0.1} taken at 20 UT on December 11 and 4 UT on
December 12. The isosurfaces viewed from the top have helical
structures that are consistent with the shape of the channel field
lines shown in Figure~\ref{3Dlines}. The height of the isosurface at
20 UT on December 12 is about 2 Mm. In fact, as the negative flux
thread emerges, the increase of the isosurface is more enhanced in the
lateral direction along the PIL than in the vertical direction. The
overlying field lines may prohibit the emerging flux from rapidly
expanding to the upper layer.

\begin{figure}[tb]
\begin{center}
    \plottwo{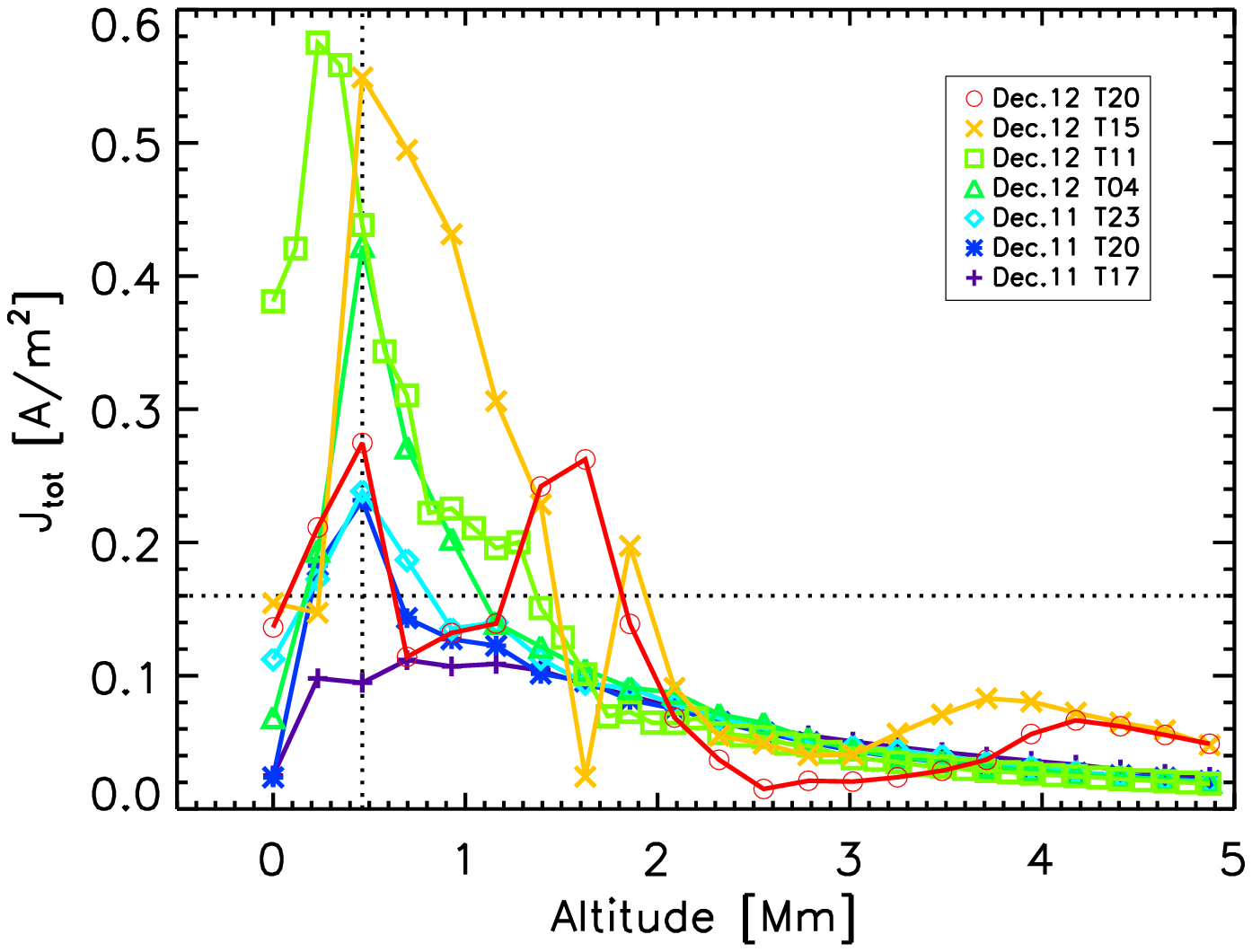}{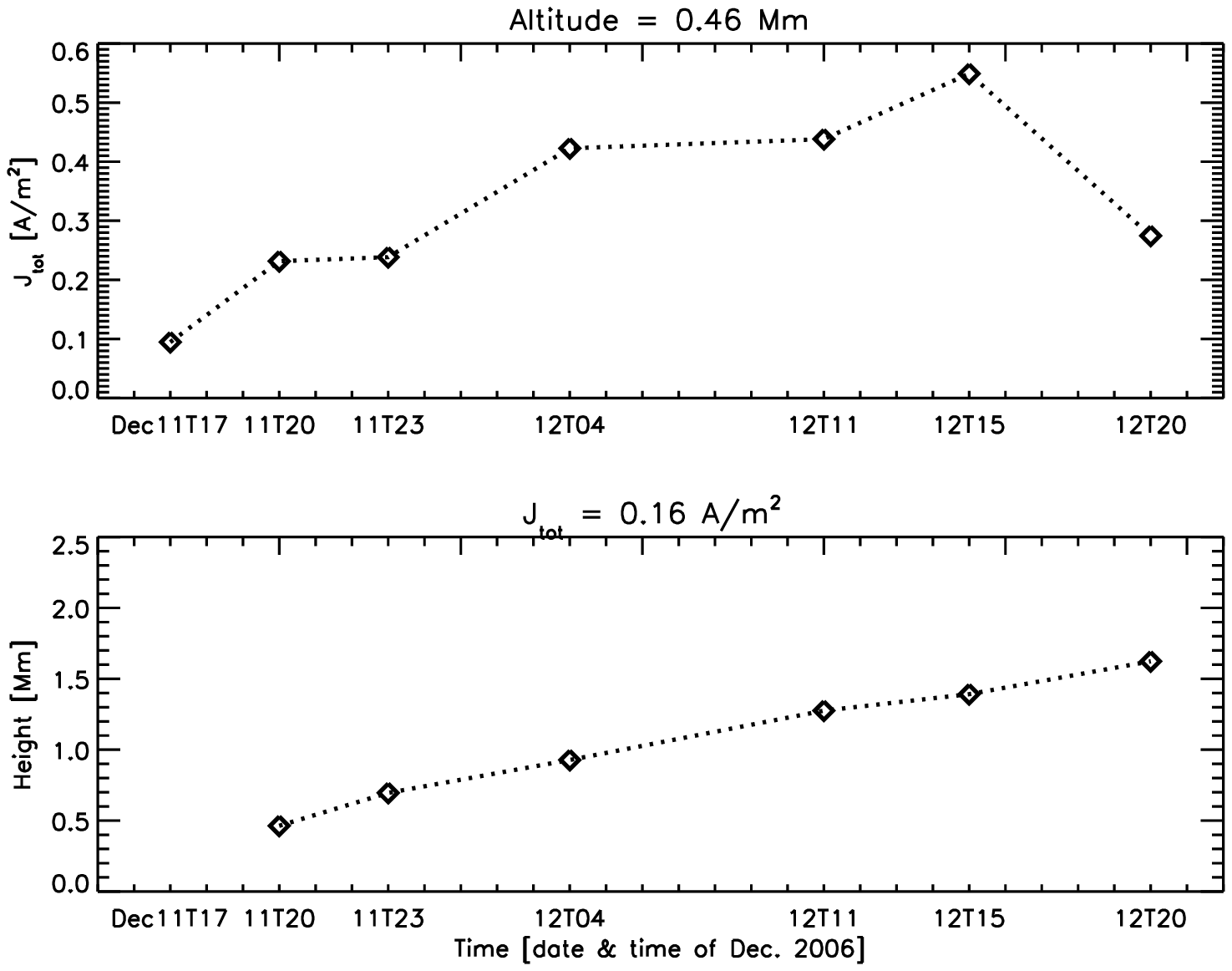}\\
    \caption{{Left} panel : total current density $|J_\textrm{tot}|=\sqrt{{J_x}^2+{J_y}^2+{J_z}^2}$
    over the altitude above the region of the negative flux emergence.
    Each value of the total current was taken from the average over $3\times3$
    pixels centered on the position of peak height-integrated current density at each altitude.
    {Right} panel : Time profiles of the total current density measured at $0.46$ Mm ({upper panel})
    and height at which the value of the total current density is \amm{0.16} ({lower panel}).}
    \label{curplot}
\end{center}
\end{figure}

The left panel of Figure~\ref{curplot} shows the spatial variations of
the total current density over the altitude at different observation
times. It is clear from the figure that high density of current is
confined to heights below about 2~Mm in all observations and the total
current density decreases to nearly zero in the higher layer. This
trend of decrease in the higher layer is not surprising considering
the general expectation that the coronal field gets more
potential-like as well as the field strength gets weaker as the
altitude increases \citep{Jing08}. What is more interesting to us is
the finding of the narrow vertical extent of the high-density current
layer since it implies that the newly emerging flux that carries
electric current is vertically confined.

The high-density current layer, even though vertically confined,
underwent temporal changes. The right upper panel of
Figure~\ref{curplot} presents the time profile of the total current
density measured at a fixed height of $0.46$ Mm and shows that the
current at this lower layer increases as the channel evolves. This
result indicates that the emerging flux thread carries electric
current into the lower corona. Moreover, the right lower panel shows
that the height of the same total current of \amm{0.16} increases by
about $1.1$ Mm in $27$ hours. This means that the electric current not
only increases at the lower layer but also transfers to the upper
layer because of the emergence of the flux thread.

\section{Summary and Discussion}

\subsection{Formation of the magnetic channel}
We have analyzed the formation process of the magnetic channel in AR
10930 in detail, focusing on the region of a negative flux thread
that comprises the magnetic channel, and obtained the following
results.
\begin{enumerate}
\item
Negative flux threads that comprise the magnetic channel have
elongated shape from the early phase of their emergence, and
correspond to penumbral filaments seen in the intensity maps.
\item
Upflows of $-0.5$ to \kms{-1.0} are detected inside the thread and
downflows of 1.5--\kms{2.0} near both tips of the thread.
\item
The magnetic fields of the channel structure are highly sheared from
the early phase of the formation and their averaged inclination
angle increases from $95^{\circ}$ ($5^{\circ}$ inclined from the
surface) to $112^{\circ}$ ($22^{\circ}$ inclined from the surface)
as the thread emerges.
\item
The map of the vertical component of electric current density displays
a pair of strong vertical current threads of opposite polarity along
the neutral line between the negative flux thread and the positive
umbra.
\item
The coronal magnetic field computed using the NLFFF extrapolation
indicates the emergence of highly sheared arch-shaped field lines inside
the slightly sheared arcades that resemble the upper part of the twisted
flux tube.
\item
The total current density in the lower corona (at the height of $0.46$
Mm) increases (by \amm{0.45} for 22 hr) and slowly transfers to the
upper layer ($\Delta$height$ =1.1$ Mm for 27 hr) due to the emergence
of the flux thread.
\end{enumerate}

These results indicate that the highly sheared magnetic channel is
formed by the subsequent emergence of the current-carrying magnetic
fields, rather than the photospheric flow after the flux emergence.
{In case the emerging field was initially potential and was sheared by
the photospheric flow, there should be observed a sufficient amount of
velocity gradient of horizontal flows across the neutral line.
Horizontal flows parallel to the neutral line but toward opposite
directions on each side of the neutral line may stretch and shear the
magnetic field. And the amount of the shear, the shear angle, would
depend on the velocity gradient across the neutral line. We measured
the mean value of the shear angles within a distance of 500 km from
the channel's neutral line and the value was about $72^{\circ}$ at 20
UT on December 12. If we assume that the semi-circular field emerged
at a constant emerging speed of $0.5$ -- \kms{1.0}, based on the
upflow speed obtained from the Dopplergram, then the magnetic field
should have been tilted at $72^{\circ}$ within $8$ -- $16$ minutes. In
this case, the required velocity difference between horizontal flows
on each side of the neutral line would be quite large, i.e., $3$ --
\kms{6}. However, we could not find such a large velocity difference
across the neutral line of the magnetic channel from the time series
of FG $V/I$ images. Horizontal flows on each side of the neutral line
were toward the same direction and the speed was comparable with the
value of around \kms{1}. Therefore, we suggest that the effect of the
photospheric flow after the flux emergence may be insignificant in the
formation of the magnetic channel.}

As a matter of fact, the picture of the formation of the magnetic
channel by the emerging magnetic flux was already pointed out by
previous studies \citep{Zir93,Kub07,wang08}. In these studies, the
pattern such as negative polarities alternate with positive ones is
schematically explained by the emergence of multiple bipoles. Although
they conjectured that these emerging bipoles are likely to be carrying
shear and twist into the corona, however, their studies were
insufficient to reveal where and how the shear or twist was supplied.
In contrast, our study investigates the very early phase of the flux
emergence and provides strong evidence that each emerging flux that
comprises the magnetic channel is actually a fine-scale twisted flux
tube.

Our study indicates that the elongated shape of the negative
magnetic flux thread constituting the magnetic channel is an
intrinsic property of its own rather than a result of either
stretching by horizontal motions \citep{Zir93} or squeezing by
surrounding magnetic flux \citep{wang08}. The flux thread
corresponds to a penumbral filament seen in the intensity maps, and
these two features appear to be two different manifestations of the
same structure: an emerging twisted flux tube. This result is in
line with the study of \citet{Rub07} who found the evidence that
supports the idea that the dark-cored penumbral filaments may be
flux tubes carrying Evershed flows.

\subsection{NLFFF extrapolation in the lower corona}
\begin{figure}[tb]
\begin{center}
    \plottwo{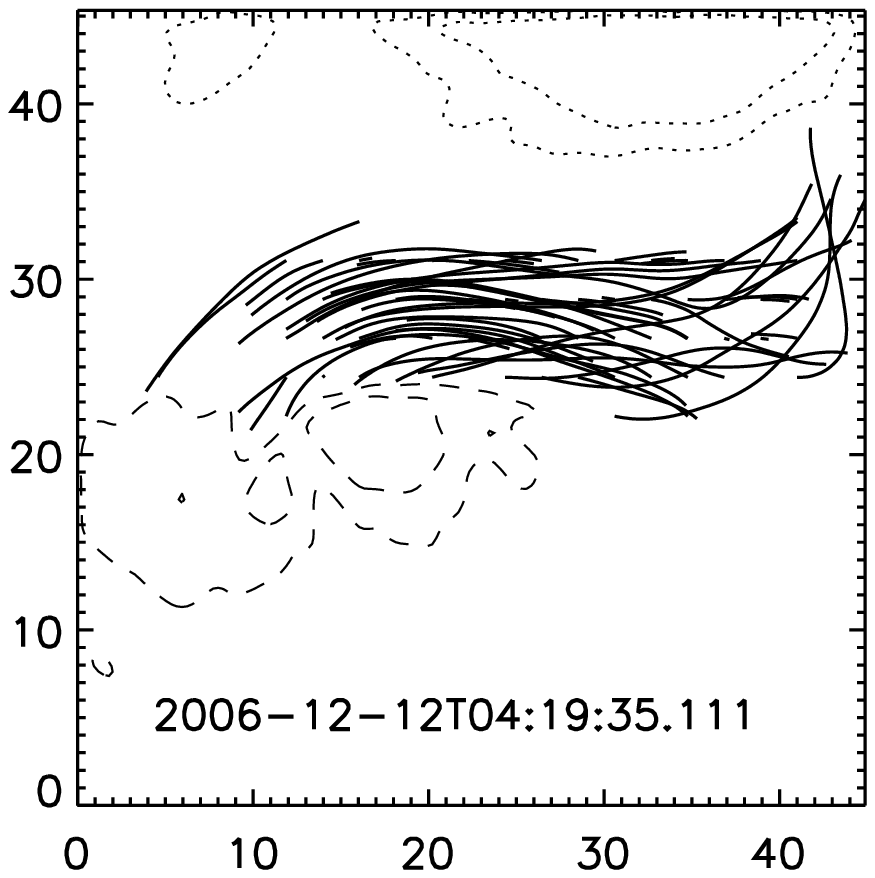}{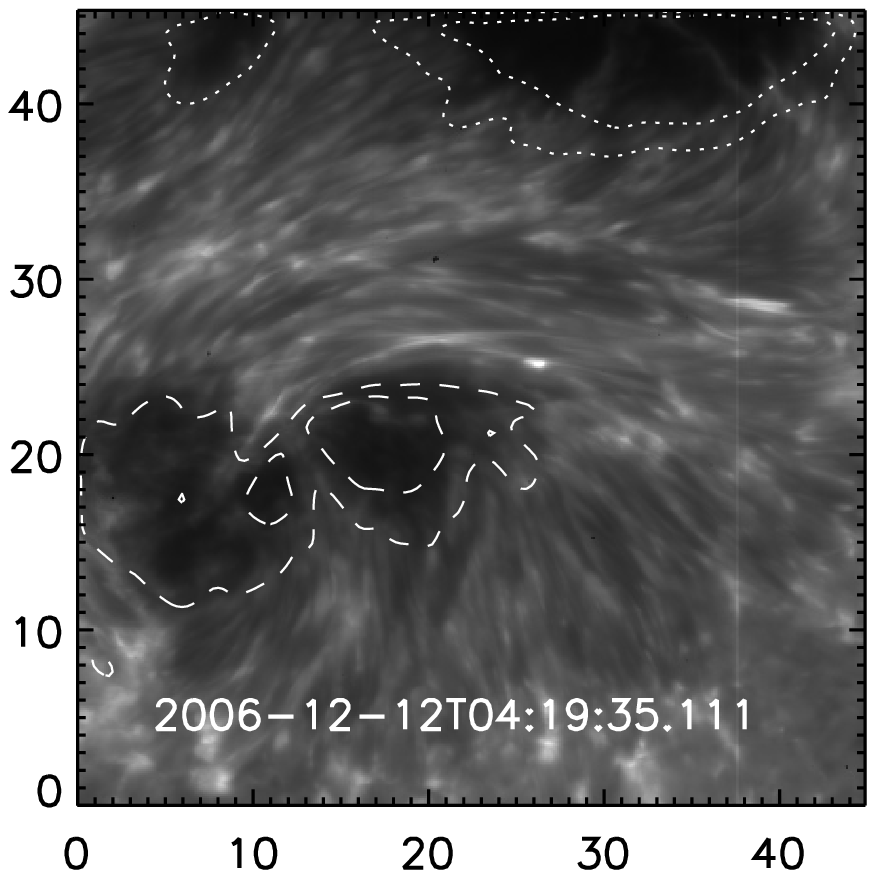}\\
    \caption{{Left} panel : magnetic field lines
    reconstructed from the NLFFF extrapolation
     of which height is less than 2~Mm.
     {Right} panel : Ca~{\footnotesize II} image with the same FOV as the
     bottom boundary condition of the NLFFF extrapolation. Dotted lines indicate
     isocontours of negative polarity and dashed lines positive
     polarity of the bottom vertical magnetogram in both panels.}
     \label{fieldlines_ca}
\end{center}
\end{figure}

The narrow width of the magnetic channel observed in the photospheric
magnetograms hints that its vertical extension may be also quite low,
comparable to the size of its width. Both reconstructed magnetic
fields (Figure~\ref{3Dlines}) and the isosurface of transverse current
density (Figure~\ref{3Dcur}) also indicate that the channel structure
lies in the lower atmosphere corresponding to chromospheric level.
Then it should be checked if it is reasonable to apply the NLFFF
extrapolation to such a low atmosphere. We have compared the
extrapolated field lines lower than 2~Mm and chromospheric structures
observed in a Ca~{\footnotesize II} image taken by SOT/\textit{Hinode}
(Figure~\ref{fieldlines_ca}). The field lines in the left panel of
Figure~\ref{fieldlines_ca} relatively well follow the topology of
chromospheric structures in the Ca~{\footnotesize II} image. Note that
the region where we reconstructed NLFFF is the center of the strong
active region. The unsigned field strength at the photosphere is over
3000~G in the positive polarity and over 2000~G in most of the
interested region. Therefore, we expect that the plasma-$\beta$ would
be less than unity in the upper chromosphere of the active region and
the force-free assumption could be reliably applied in such a low
portion. The plasma-$\beta$ model derived by \citet{Gary01} also shows
that $\beta < 1$ at around 2~Mm above an active region although its
value depends on both the magnetic and the pressure model they chose.

\subsection{Relationship with flares}
Is the formation of the magnetic channel important in the occurrence
of flares? Our finding that strong current flows along the flux thread
that comprises the magnetic channel suggests a positive answer to this
question. It has been frequently pointed out that the flux emergence
already carrying currents and helicity in the active region helps to
add magnetic free energy into the region \citep{Ish98,Bro03}. Adding
magnetic free energy could destabilize the pre-existing magnetic field
in active region so that it may be subject to more eruptive events
such as flares, coronal mass ejections and filament eruptions.
Recently, increasing evidence is suggested from observations that
support the idea of emerging currents associated with the emerging
twisted magnetic flux from below the photosphere
\citep{Kur87,Tan91,Lek96,Ish98,Sch08}.

Our another finding that the electric current is confined to below 2
Mm above the surface, however, seems to make a problem. This height is
too low for the expected height of the reconnection that produces the
flare. Based on the standard flare model and considering the size and
distance of the two-ribbon footpoints, the reconnection in charge of
major flares should occur at the height of around 10 Mm, which is much
higher than 2 Mm. One probable solution to this problem is to suppose
that the strong electric current is gradually transferred to the
higher layer, as is implied from our results from NLFFF models and
would form the current sheet there, an essential component of the
reconnection. Moreover, magnetic reconnection is likely to occur
between newly emerging flux and pre-existing ambient field and can
also play a role in transferring the free magnetic energy carried by
the emerging flux tube to the overlying coronal field. Since the
emergence of the flux thread, brightening was continuously detected
along the flux thread in the Ca~{\footnotesize II} images taken by
SOT/\textit{Hinode}. This brightening may represent subsequent
magnetic reconnection of this kind, through which we may relate the
emerging flux tube of much smaller scale to the December 13 flare with
much larger size. Summing up, it seems that the formation of magnetic
channel in AR 10930 plays a role in the December 13 flare in that it
destabilizes the ambient magnetic field by carrying free energy into
the corona.

\subsection{Fine structure of AR magnetic fields}
This study on the flux thread gives us implications not only on the
structure of magnetic channel itself but also on the fine structure of
an active region magnetic field. In the case of AR 10930,
\citet{Sch08} and \citet{Mag08} already suggested the emergence of a
twisted flux tube by retrieving coronal magnetic fields using the
NLFFF extrapolation and by checking the temporal variation of the
shear angle along the PIL, respectively. Compared to the scale size of
their interests, the channel structure we focused on is a lot smaller,
a few Mm in height. Then how could it be explained that twisted flux
tubes in different scale size coexist in the same region? It seems
that the magnetic fields of channel structure may be a sub-structure
of larger scaled flux tube. The simple model of an active region
commonly adopted is that the whole active region is a twisted flux
tube. This assumption frequently well describes the large-scale
structure of an active region such as sigmoid in coronal observations.
However, as we observe active regions with higher spatial resolution,
we find more complex and fragmentary magnetic structures, such as
magnetic channel in our case. It gives us an impression that the
active region as a whole twisted flux tube consists of a number of
segments of twisted flux ropes. The successive emergence of
small-scale flux often has been explained by the emergence of
undulating single or multiple flux ropes
\citep{Ish98,Low01,Kur02,Sch09}. The formation of active region
filaments is also explained by emergence of twisted flux tube by some
authors \citep{Lit95,Oka08}. Depending on the size of these segments
of twisted flux rope, different features would be observed at the
photosphere.

\acknowledgments We thank the referee for constructive comments that
significantly improved the manuscript, and Jeongwoo Lee for helpful
comments. This work was supported by the Korea Research Foundation
Grant funded by the Korean Government (KRF-2008-220-C00022). Ju Jing
was supported by NSF under grants ATM 09-36665 and ATM 07-16950, and
Haimin Wang by US NASA NNX-09AQ90G. Thomas Wiegelmann was supported by
DLR-grant 50 OC 0501. \textit{Hinode} is a Japanese mission developed
and launched by ISAS/JAXA, with NAOJ as domestic partner and NASA and
STFC (UK) as international partners. It is operated by these agencies
in cooperation with ESA and NSC (Norway).

\end{document}